\documentclass[aps,prf,10pt,superscriptaddress,showkeys,longbibliography]{revtex4-1}
\usepackage{amsmath}
\usepackage{graphicx}% Include figure files
\usepackage{dcolumn}% Align table columns on decimal point
\usepackage{bm}% bold math
\usepackage{siunitx}
\usepackage{xcolor}
\newcommand{\new}[1]{{\color{black} #1}}

\begin{document}

\title{Collective Effects in Breath Figures}

\author{Ambre Bouillant}%\aff{1}, 
 \affiliation{Physics of Fluids Group, Faculty of Science and Technology, University of Twente, 7500 AE Enschede, The Netherlands}
\affiliation{Laboratoire Matière et Systèmes Complexes, UMR 7057 CNRS, Université Paris Cité, 10 rue A. Domon et L. Duquet, 75013, Paris, France}
\author{Jacco H Snoeijer}%\aff{1},
\affiliation{Physics of Fluids Group, Faculty of Science and Technology, University of Twente, 7500 AE Enschede, The Netherlands}
 \author{Bruno Andreotti}%\aff{2} Decide order and diplomatic authors}
\affiliation{Laboratoire de Physique de l'ENS, UMR 8550 CNRS, Ecole Normale Sup\'erieure, Universit{\'e} Paris Cit\'e -- Sorbonne~Universit{\'e}, 24 rue Lhomond, 75005 Paris, France}

\date{\today}

\begin{abstract}
Breath figures are the complex patterns that form when water vapor condenses into liquid droplets on a surface. The primary question concerning breath figures is how the condensing vapor is allocated between the growth of existing droplets and the nucleation of new ones. Although numerous theoretical studies have concentrated on scenarios resulting in highly polydisperse droplet ensembles, a companion paper [Bouillant {\em et al.}, submitted] demonstrates that nearly monodisperse patterns can be achieved on defect-free substrates in a diffusion-controlled regime. The objective of this work is to present a theoretical framework that elucidates the formation and evolution of nearly-monodisperse patterns in breath figures. We discover that, following a short nucleation phase, the number of droplets remains constant over an extensive range of timescales due to collective effects mediated by the diffusion of vapor. The spatial extent of these diffusive interactions is identified through asymptotic matching, based on which we provide an accurate description of breath figures through a mean-field model. The model accounts for the sub-diffusive growth of droplets as well as for the arrest of nucleating new droplets, and reveal the scaling laws for the droplet density observed in experiments. Finally, droplets expand and ultimately coalesce, which is shown to trigger a scale-free coarsening of the breath figures.
\end{abstract}
%\keywords{}

\maketitle

\section{Introduction} %
``The manner in which aqueous vapor condenses upon ordinarily clean surfaces of glass or metal is familiar to all" \cite{Rayleigh1911}. Yet, the discussion of breath figures, as initiated by Rayleigh and Atkin in 1911 \cite{Rayleigh1911}, still continues today and holds many unresolved puzzles. Breath figures are the intricate patterns that form when water vapor condenses into liquid droplets on a surface \cite{baker1922lxv}. The process of nucleation, in which nascent structures of a stable phase emerge in a metastable phase, is crucial to the formation of breath figures \cite{kashchiev2000nucleation,beysens2022physics}. In this context, the nucleation of nanometric droplets occurs on the solid through a heterogeneous mechanism that is highly sensitive to the substrate nature \cite{Knobler1991}. On rough, patterned or chemically heterogeneous solids, surface defects can lower the energy barrier to nucleation, resulting in breath figures that reflect the substrate heterogeneity, such as the density of defects. In contrast, on atomically smooth substrates, nucleation is stochastic and only depends on the energy barrier associated with the formation of a critical nucleus, reflecting the substrate energy \cite{Varanasi2009,Sikarwar2011,Lopez1993,Enright2012}. The substrate properties can also affect the droplets' growth through the influence of the substrate wettability on the vapor flux absorbed by the droplet \cite{Gelderblom2011,Popov2005}. 
Secondary nucleation events can also be impacted by air-mediated interactions between droplets. 
Already formed droplets tend to inhibit new droplet formation in their vicinity, creating dry Regions of Inhibited Condensation (RIC). These RICs have been studied around hygroscopic materials such as ice \citep{nath2018duelling} or saltwater \citep{guadarrama2014droplet}, as well as around hydrophilic coatings \citep{schafle2003subpattern,yu2021droplet} (see also \cite{beysens2022physics}). Additionally, the latent heat released during condensation can trigger convection above a diffusive thermal boundary layer, mixing vapor above the droplets \citep{Beysens2006}. When droplets are sufficiently close together, many studies \citep{Briscoe1991,Sokuler2010interspacing,guadarrama2014droplet,picknett1977evaporation} assume that they behave collectively as a homogeneous liquid film, with a thickness equal to the volume of liquid deposited per unit area.

At a more advanced stage of the condensation process, the occurrence of coalescence events between droplets abruptly changes drop configurations, resulting in intermittent dynamics and increased polydispersity in the breath figure pattern, ultimately promoting gravity-driven droplet shedding \cite{Bintein2019,Trosseille2019}. On rough solids, fusions between neighboring droplets free up the surface defects, enabling secondary nucleation events and markedly increasing the pattern polydispersity. In recent decades, experiments performed on clean surfaces have suggested the emergence of a universal scaling form of the droplet size distribution evolution \cite{Viovy1988scaling,Blaschke2012,Stricker2022universality,haderbache1998numerical}. However, the underlying mechanisms responsible for the formation and evolution of breath figures are not yet fully understood  \citep{beysens2022physics}. 

Condensation is a natural phenomenon that has been harnessed for various applications, from providing a valuable source of water in deserts for humans \cite{nikolayev1996water,liu2022water} and animals \cite{parker2001water,munne1999role,hill2015role}, to heat exchangers for cooling systems \cite{bortolin2022heat}, desalination systems \cite{khawaji2008advances}, designing nano-emulsions \cite{Guha2017} with innovative optical properties \cite{goodling2019colouration}, and patterning substrates for microfabrication \cite{boker2004hierarchical,zhang2015breath}. In order to optimize 
the gravity-driven droplet shedding in these applications, it is crucial to control the density, size, and polydispersity of dew patterns. This requires a fine control of the substrate heterogeneity, such as the presence of defects that promote nucleation, and properties. Strategies for controlling dew patterns may involve surface patterning \cite{Park2016,Trosseille2019,Zhao1995,Varanasi2009,Bintein2019} or electrowetting \cite{Baratian2018}. Therefore, understanding the formation and evolution of breath figures is essential for predicting and controlling dew patterns in a wide range of applications.

In this study, we investigate the early stages of breath figure formation on defect-free surfaces and the ensuing pattern coarsening due to drop merging. The work provides the interpretation-framework for an experimental companion paper \cite{PRL_companion}, which shows the emergence of nearly monodisperse breath figures and their subsequent coarsening. Here we develop a many-droplet theory for nucleation and growth, based on an asymptotic matching between the vicinity of drops and a far field, which describes droplet interactions through the quasi-static diffusion of the vapor phase. Our model reveals two distinct phases of growth. At short times, the droplets grow diffusively while the the number of drops increases linearly in time. However, at longer times, we show that, in the absence of turbulent mixing in the vapor phase, nucleated droplets deplete the humidity in their vicinity. This depletion creates a boundary layer that grows as the pattern evolves. As droplets approach each other, interactions mediated through the surrounding air screen the humidity, resulting in a lower effective humidity experienced by the droplets. This humidity screening effect leads to a sub-diffusive droplet growth, and also affects the energy barrier to nucleate new droplets in a highly sensitive manner -- ultimately leading to the arrest of nucleation. A central question that is addressed in this paper is how the experimentally observed number of drops (starting around the micron-scale \cite{PRL_companion}) is related to the actual nucleation rate (at the nano-scale). Finally, we address the ultimate regime of the nearly-monodisperse breath figures, where drop come to contact and exhibit coarsening by coalescence. It is demonstrated that the coarsening exhibits a scale-free dynamics, explaining the scaling laws observed in experiment \cite{PRL_companion}.

\section{Drop nucleation}
\label{sec:SingleDropDescription}

\begin{figure}
\centering
\includegraphics[width=\textwidth]{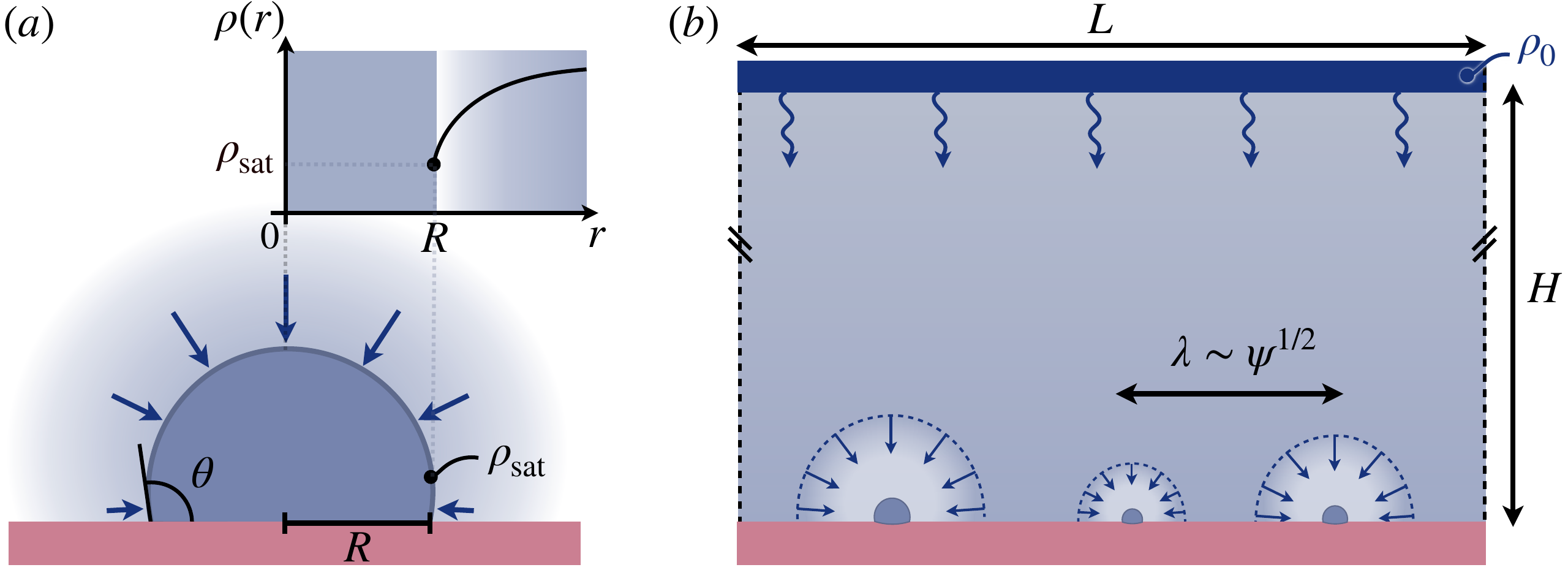}
\caption{Nucleation and growth of breath figures. (a) Isolated drop of base radius $R$ wetting its substrate with a contact angle $\theta$. The arrows indicate the local vapor flux driving the droplet growth. The inset shows a typical vapor density $\rho$ along the substrate, increasing from the saturated value $\rho_{\rm sat}$ at the droplet's interface. (b) On the scale of the cell (height $H$), the drops act as ``sinks" of vapor, which modify the humidity fields in their vicinity. In most of the paper, we focus on the regime where typical distances between the drops $\lambda$ are large compared to the typical drop size, i.e. in a regime prior to drop coalescence. This gives a hierarchy of scales $R \ll \lambda \ll H$ that enables a solution of the many-droplet diffusion problem by matched asymptotics. In the calculation, we use periodic boundary conditions in the horizontal direction, so that the cell size is $L\times L \times H$.}
\label{fig:FarAdNearFields}
\end{figure}

In this study, we employ the Classical Nucleation Theory \cite{kashchiev2000nucleation,kelton2010nucleation} to describe the nucleation of drops on a defect-free rigid substrate as in \cite{beysens2022physics}. The drops under consideration are expected to be nanometric in size, allowing us to disregard the effects of gravity. Consequently, the liquid assumes the shape of a spherical cap, the geometry of which is characterized using the horizontal base radius $R$ and a contact angle $\theta$, as depicted in Fig.~\ref{fig:FarAdNearFields}(a). At equilibrium, the contact angle is selected by the surface energies of the solid-vapor interface ($\gamma_{sv}$), the solid-liquid interface ($\gamma_{s\ell}$) and the liquid-vapor interface ($\gamma$) through Young's law: $\gamma \cos \theta = \gamma_{sv}-\gamma_{s\ell}$. The droplet volume can be written as $V = \frac{\pi}{3}R^3\mathcal{E}(\theta)$ where the function $ \mathcal{E}(\theta)\equiv {(2+\cos\theta)(1-\cos\theta)^2}/{ \sin^3\theta}$ accounts for the influence of the contact angle. For a hemispherical drop one verifies that $\mathcal{E}(\pi/2) = 2$, while at small $\theta$ we find $\mathcal{E}(\theta)\sim 3\theta/4$. The number of molecules $n$ contained within the drop is then defined as the ratio between its volume $V$ and the volume of a single molecule $m \big/\rho_\ell$, where $\rho_{\ell}$ and $m$ are the liquid mass density and the mass of a single molecule. We thus find
\begin{equation}
%n =\frac{\rho_\ell V }{m} = \frac{\pi \rho_\ell}{3 m}R^3\mathcal{E}(\theta)
n =\frac{ V }{m/\rho_\ell} = \frac{\pi \rho_\ell}{3 m}R^3\mathcal{E}(\theta).
 \label{eq:MicroMacroCorrespondence}
\end{equation}

\new{We now consider the nucleation of a droplet containing $n$ molecules at the surface of the solid substrate, from a supersaturated vapor phase. The creation of new interfaces (liquid-vapor, solid-liquid and solid-vapor) associated to the creation of the drop comes at an energetic cost. This energy penalty scales with the surface $\sim R^2 \sim n^{2/3}$. The nucleation is still facilitated by the energetic gain per molecule, associated to phase change from the vapor to liquid. The relevant thermodynamic potential $\Phi$ is referred to as the external Gibbs free energy, exergy, available work potential, or free energy, depending on the community \cite{eslami2011thermodynamic,beysens2022physics,diu2007thermodynamique}. %\textcolor{red}{cite Beysens, the french textbook + the paper suggested by the referee}. 
As is recalled in Appendix \ref{app:ExternalG}, the change in $\Delta \Phi$ due to the phase change involves the chemical potential of the liquid ($\mu_\ell$) and  vapor phases ($\mu_v$) evaluated at the pressure $p_v$ and temperature $T$ of the vapor phase. Specifically, the contribution to $\Delta \Phi$ per molecule reads
\begin{equation}
\Delta \mu = \mu_\ell(p_v,T)-\mu_v(p_v,T) =- \Lambda k_BT, \quad {\rm with} \quad \Lambda \simeq \ln\left( \rho \big/ \rho_{\mathrm{sat}} \right),
\end{equation}
where we assumed the liquid to be incompressible and the vapor to behave as an ideal gas. In this expression, $\rho$ is the vapor mass density and $\rho_{\mathrm{sat}}$ its saturated value at a given temperature $T$. The expression of $\Lambda$ given in Appendix A is here simplified assuming that the volume per molecule in the liquid phase is assumed to be much smaller than in the vapor phase. Combined with the interfacial energies, working out the geometric details,} one can write the total change in the thermodynamic potential as \cite{kashchiev2000nucleation,beysens2022physics}
\begin{eqnarray}
\frac{\Delta \Phi}{k_B T} &=& \frac{3}{2} \left(2\chi n^2 \right)^{1/3} - n \Lambda, 
\label{eq:TotalEnergyVariations}
\end{eqnarray}
where we introduced $\chi$, 
\begin{equation}
 \chi \equiv \dfrac{4\pi}{3} \left(\dfrac{ \gamma}{k_B T} \right)^3 \left(\dfrac{m}{ \rho_\ell } \right)^2 \mathcal{E}\sin^3 \theta.
\end{equation}
This dimensionless parameter isolates the droplet's geometrical influence to the energy barrier. Equation (\ref{eq:TotalEnergyVariations}) contains the Gibbs-Kelvin effect, which combines capillary effects and the chemical potential dependence on vapor partial pressure. Importantly, $\mu_\ell$ and $\mu_v$ are evaluated at the pressure and temperature of the vapor phase, as shown in the Appendix \ref{app:ExternalG}. The model yielding equation~\ref{eq:TotalEnergyVariations} must be refined in the limit of small nuclei (with only a few molecules) by incorporating more complicated dependences of some parameters \cite{wu1996nucleation}.

The thermodynamic potential $\Delta \Phi$ thus initially increases with droplet size $n$, with a scaling $\sim n^{2/3}$ owing to surface effects, while the gain in energy due to the actual condensation $\sim n$. The competition between these effects gives rise to an energy barrier for nucleation $\Delta \Phi^*$, and a critical nucleus size $n^*$, corresponding to the maximum of $\Delta \Phi(n)$. We thus find:
\begin{equation}
n^*= \frac{2\chi}{\Lambda^3 },\quad{\rm and} \quad \frac{\Delta \Phi^*}{k_B T} = \frac{\chi}{\Lambda^2}.
\label{eq:energyGap}
\end{equation}
The critical nucleus and the corresponding energy barrier are thus controlled by the two dimensionless parameters: $\chi$, capturing the droplet geometrical influence to the energy barrier, and $\Lambda$, accounting for the humidity influence on the energy barrier. In other contexts, the relation between $n^*$ and $\rho/\rho_{\rm sat}$ is named Ostwald–Freundlich equation.\\

Finally, we assume the nucleation of drops to be an activated process that is described by Classical Nucleation Theory \cite{kashchiev2000nucleation}. We refer to Appendix.~\ref{app:CNT} for details. Within this framework, the number of drops nucleated per unit surface and unit time takes the form:
\begin{equation}
J = {\mathcal J} \left(\frac{\rho}{\rho_{\rm sat}}\right)^2 \exp\left(- \frac{\Delta \Phi^*}{k_B T}\right) = {\mathcal J} \left(\frac{\rho}{\rho_{\rm sat}}\right)^2 \exp\left(- \frac{\chi}{\Lambda^2}\right).
\label{eq:rateenergyGap_final}
\end{equation}
In this expression $\mathcal J$ can be interpreted as an attempt frequency per molecule, while the factor $(\rho/\rho_{\rm sat})^2$ reflects the fact that the overall collision frequency is quadratic in vapor density $\rho$ (cf. Appendix~\ref{app:CNT}). Most importantly, however, the nucleation rate involves a Boltzmann factor that is controlled by the energetic barrier.

\section{Collective effects during growth}
\label{sec:MicroscopicDescription}

The nucleation described above is a stochastic process, which depends on the local humidity at the substrate. Once drops are formed, however, they grow by further absorbing vapor. This growth is typically limited by the (deterministic) diffusive transport of vapor inside the cell towards the drops. This diffusive transport is described in this section, where we consistently account for collective effects due to droplet interactions.

\subsection{Theoretical set-up}

We model the growth of droplets by the quasi-steady diffusion of vapor, i.e. by solving $\nabla^2 \rho = 0$. This assumes that the transient time for the evolution of the vapor concentration is much smaller than the time-scales for the growth of the drops. So, we consider that the vapor field to be instantaneous, in the sense that they are much smaller than the time-scales of the evolution of the number of drops. The complexity of the problem arises from the intricate, mixed boundary conditions that slowly evolve over time as the breath figure evolves, and which need to be found self-consistently. At a liquid-vapor interface, the air is saturated with vapor, so that $\rho = \rho_{\rm sat}$ is imposed at the surface of the drops. Between the drops, in the dry regions of the substrate, the normal component of the flux $\nabla \rho \cdot \mathbf n$ must vanish, as molecules cannot penetrate the substrate. These boundary conditions are illustrated on the level of a single, isolated drop in Fig.~\ref{fig:FarAdNearFields}a, for which analytical solutions are available \citep{Popov2005}. The breath figure is driven by imposing a humidity $\rho=\rho_0$ at the top plate at $z=H$ (Fig.~\ref{fig:FarAdNearFields}b), which leads to condensation when $\rho_0 > \rho_{\rm sat}$. We introduce the relative humidity $\mathcal R={\rho_0}\big/{\rho_{\rm sat}}$, which is taken larger than unity throughout. 

\begin{figure*}[t]
\centering
\includegraphics[width=\textwidth]{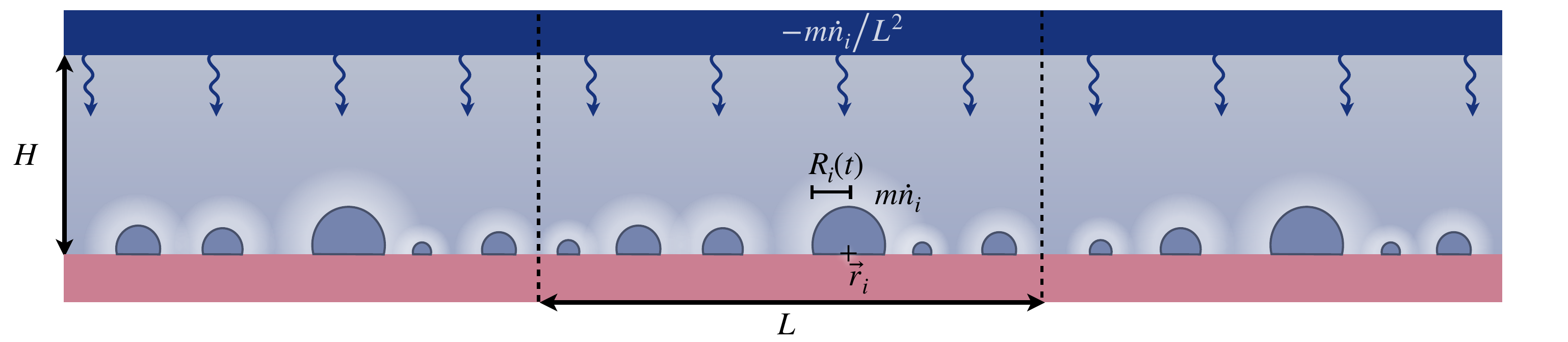}
\caption{Modeling the vapor flux collected into dew droplets with the electrostatic analogy. A homogeneous surface ``charge" $- m\dot n_i/L^2$ imposed at the ceiling ($z=H$) provides a vapor flux that distributes over the $n_i$ droplets (modelled as point sinks), each absorbing a ``charge" $m\dot n_i$.}
\label{fig:Schemanip}
\end{figure*}

\subsection{Non-interacting drops}
The quasi-steady diffusive growth of an isolated sessile drop, spherical cap of contact angle $\theta$ and base radius $R$ as sketched in Fig.~\ref{fig:FarAdNearFields}(a), is well understood. 
Introducing $n_i$ as the number of molecules inside a droplet labelled $i$, the growth law is given by \citep{Popov2005}
\begin{equation}
m \dot n_i = 2 \pi \mathcal{F}(\theta) R_i D\left(\rho_{\infty}^i-\rho_{\rm sat}\right),
% \mathcal{F}(\theta)&=&\frac{\sin \theta}{2(1+\cos \theta)}+2 \int_{0}^{\infty} \frac{1+\cosh 2 \theta \tau}{\sinh 2 \pi \tau} \tanh [(\pi-\theta) \tau] d \tau.\nonumber
\label{eq:GrowRateHydrodynamicRegime}
\end{equation}
where $D$ is the diffusion coefficient of vapor molecules in air, while
\begin{eqnarray}
 \mathcal{F}(\theta)&=&\frac{\sin \theta}{2(1+\cos \theta)}+2 \int_{0}^{\infty} \frac{1+\cosh 2 \theta \tau}{\sinh 2 \pi \tau} \tanh [(\pi-\theta) \tau] d \tau.\nonumber
\end{eqnarray}
the function $\mathcal F(\theta)$ follows from solving Laplace's equation $\nabla^2 \rho=0$, with boundary conditions $\rho=\rho_{\rm sat}$ on the surface of the spherical cap, a no-flux condition at the dry substrate, while $\rho_\infty^i$ is the humidity at infinity \citep{Popov2005}. $\mathcal{F}(\theta)$ tends to $2/\pi$ as $\theta$ tends to $0$ and is equal to $1$ for the hemispherical case ($\theta=90^{\circ}$). The diffusive law has been validated experimentally \citep{Gelderblom2011} for an isolated evaporating drops with $\rho_\infty^i$ taken constant. \\
For later reference, we will express the growth law for $\dot n_i$ in terms of $\dot R_i$. Making use of (\ref{eq:MicroMacroCorrespondence}), we can write (\ref{eq:GrowRateHydrodynamicRegime}) as
\begin{equation}
\frac{d(R^2_i)}{dt}= \mathcal D \;\left( \frac{\rho_{\infty}^i}{\rho_{\rm sat}}-1\right), 
%\quad{\rm with}\quad \mathcal D=\frac{4\mathcal{F}(\theta)\rho_{\rm sat}}{\mathcal E \rho_\ell } D,
\label{eq:growthRsquared}
\end{equation}
where we introduced the effective diffusion coefficient $\mathcal D$ as
\begin{equation}
 \mathcal D=\frac{4\mathcal{F}\rho_{\rm sat}}{\mathcal E \rho_\ell } D.
 \label{eq:mathcalD}
\end{equation}
Dropping the label $i$, we see that isolated drops grow as $R = \sqrt{(\mathcal R - 1)\mathcal D (t-T)}$, where $T$ is the time of creation of the drop and $\mathcal R = \rho_{\infty}/\rho_{\rm sat}$ the relative humidity at infinity. \\
The complete solution $\rho(\mathbf r)$ around a spherical cap is intricate, but in the limit $|\mathbf r - \mathbf r_i| \gg R_i$ it reduces to \citep{Popov2005}, 
\begin{equation}\label{eq:inner_asymptotics}
\rho (\mathbf r - \mathbf r_i) \simeq \rho_{\infty}^i + (\rho_{\rm sat}-\rho_{\infty}^i)\mathcal{F}\frac{ R_i}{|\mathbf r - \mathbf r_i|}. 
\end{equation}
In the far-field, the detailed geometry of the droplet is thus no longer reflected in the spatial structure of the field: the asymptotic solution corresponds to an isotropic sink at position $\mathbf r_i$. The growth law \eqref{eq:GrowRateHydrodynamicRegime} is indeed recovered from the integration of the diffusive flux $ - D \nabla \rho$ obtained from \eqref{eq:inner_asymptotics}, over the half-spherical surface $2\pi |\mathbf r - \mathbf r_i|^2$. It is instructive to write the far-field solution directly in terms of $\dot n$, 
\begin{equation}\label{eq:inner_asymptotics_bis}
\rho (\mathbf r - \mathbf r_i) \simeq \rho_{\infty}^i - \frac{m \dot n_i}{2\pi D |\mathbf r - \mathbf r_i|}.
\end{equation}
This form gives an explicit relation between the strength of the sink and the growth rate $\dot n_i$ of the drop.

\subsection{Interacting drops}
In the presence of many drops, one expects the same growth law, as long as the drops are sufficiently separated. Specifically, \eqref{eq:inner_asymptotics_bis} will be valid at large distance from the drop yet remaining far from neighbouring drops, \textit{i.e.} for the hierarchy of scales
\begin{equation}
R_i \ll |\mathbf r - \mathbf r_i | \ll \lambda = \psi^{-1/2}. 
\end{equation}
Here we introduced the number of drops per unit area $\psi$. The fact that drops are small, however, does not imply that drops do not interact. We will see that $\rho_\infty^i$ represents the ``effective humidity" experienced by droplet $i$, and that its value results from the presence of the ensemble of drops. Clearly, this effective humidity will in general be lower than the humidity $\rho_0$ imposed at the top-surface of the cell, since the neighbouring drops will have a screening effect. By \eqref{eq:GrowRateHydrodynamicRegime}, it is clear that the growth rates $\dot n_i$ for each of the droplets crucially relies on the value of $\rho_\infty^i$. \\
We now set out to formulate and solve the intricate many-body quasi-steady diffusion problem, $\nabla^2 \rho=0$ with appropriate boundary conditions, which enables us to compute the $\rho_\infty^i$. As sketched in Fig.~\ref{fig:FarAdNearFields}(b), we consider a system of height $H$, and length and width $L$ with periodic boundary conditions in both horizontal directions. A number $\psi L^2$ drops are placed on the substrate ($z=0$), each characterised by a position $\mathbf r_i$, base radius $R_i$ and number of molecules $n_i$. The condensation is driven by prescribing an average concentration $\rho_0 = \int_{z=H} \rho \, d^2 x$ at the top of the cell ($z=H$). As previously mentioned, the bottom substrate is subjected to a no-flux condition $\nabla \rho \cdot \mathbf n=0$, while $\rho=\rho_{\rm sat}$ at the surface of each droplet.

\subsubsection{Solution by matched asymptotics}
We solve for the field $\rho(\mathbf r)$ and the effective humidities $\rho_\infty^i$ using a matched asymptotics analysis that employs the scale separation $R_i \ll \psi^{-1/2}$, \textit{i.e.} that the typical distance between drops is large compared to their typical size. The inner problem is determined by the scale $R_i$ and corresponds to an isolated drop [Fig.~\ref{fig:FarAdNearFields}(a)]. The far-field asymptotics of this inner solution, \textit{i.e.} for $R_i \ll |\mathbf r - \mathbf r_i |$, is already given by \eqref{eq:inner_asymptotics_bis}.
The outer problem is governed by the scale $\psi^{-1/2} \gg R_i$ [Fig.~\ref{fig:FarAdNearFields}(b)]. In the outer problem, the droplets can be represented by isotropic point sinks with a singularity of the form $\rho \simeq -\dot m n_i/2\pi D |\mathbf r - \mathbf r_i|$. This makes the problem analogous to point charges in electrostatics, the charge here being directly related to the growth rate $\dot n_i$. This electrostatic analogy is sketched in Fig.~\ref{fig:Schemanip}. Besides the charges associated to the growing droplets, we impose a homogeneous surface charge density $- m\dot n_i/L^2$ at $z=H$. The total charge in the system is thereby zero, which ensures the conservation of mass: the flux imposed at the ceiling feeds the growth of the drops. Using the superposition principle, the outer solution can be written as
\begin{equation}
\rho(\mathbf r)= \rho_0 - \sum_{j=1}^{ N} \frac{m \dot n_j }{2\pi D}\mathcal{G}(\mathbf r-\mathbf r_j),
\label{eq:outer_asymptotics}
\end{equation}
where the sum runs over each drop. Here, $\mathcal G$ is the Green's function that captures not only the effect of each sink, but also of the uniform flux that feeds the drop from the ceiling, the no-flux boundary condition at $z=0$, and self-interaction through the (horizontal) periodic boundary conditions. The gauge of the Green's function is chosen such that it has a vanishing spatial average at $z=H$. With this gauge, $\rho_0$ in \eqref{eq:outer_asymptotics} represents the average humidity at the ceiling of the cell, which we will use as a control parameter for the condensation process. The explicit computation of the Green's function is provided in Appendix~\ref{app:electrostatics}. \\
The asymptotic matching consists of equating the inner expansion (\ref{eq:inner_asymptotics_bis}) far from the drop, to the outer solution (\ref{eq:outer_asymptotics}) expanded in the limit $\mathbf r \to \mathbf r_i$. 
The Green's function ${\mathcal G}(\mathbf r)$ is decomposed into its $1/|\mathbf r|$ singularity and the subdominant rest, $\tilde{\mathcal{G}}(\mathbf r)$, which is regular at $\mathbf r=\mathbf 0$. Hence we write ${\mathcal G}(\mathbf r) =\tilde{\mathcal{G}}(\mathbf r)+1/|\mathbf r|$, so that in the limit $\mathbf r \to \mathbf r_i$, 

\begin{equation}
\rho(\mathbf r) = - \frac{\dot n_i}{2\pi D |\mathbf r-\mathbf r_i|} + \rho_0 - \frac{\dot n_i}{2\pi D}\tilde{\mathcal{G}}(\mathbf 0) - \sum_{j\neq i} \frac{\dot n_j }{2\pi D}\mathcal{G}(\mathbf r_i-\mathbf r_j) + \mathcal O\left(| \mathbf r_i-\mathbf r_j |\right).
\label{eq:totalphiTopUniform}
\end{equation}
This result needs to be compared to the inner asymptotics (\ref{eq:inner_asymptotics_bis}). One observes that the leading order $1/|\mathbf r - \mathbf r_i|$ already matches, which reflects that we have chosen the appropriate strengths for the point sinks in the outer problem. However, matching the next order $|\mathbf r - \mathbf r_i|^0$, we obtain a nontrivial condition
\begin{equation}
\rho_{\infty}^i = \rho_0 - \frac{m \dot n_i}{2\pi D}\tilde{\mathcal{G}}(\mathbf 0) - \sum_{j\neq i} \frac{m \dot n_j }{2\pi D}\mathcal{G}(\mathbf r_i-\mathbf r_j).
\label{eq:totalphiTopUniform}
\end{equation}
Hence, we can express the effective humidity $\rho_\infty^i$ of droplet $i$ in terms of the growth rates $\dot n_j$ of the full ensemble of the drops; this reflects the collective nature of the problem. Having solved for the effective humidities, we can return to \eqref{eq:GrowRateHydrodynamicRegime} and write an equation that only contains the growth rates:
\begin{equation}
 \frac{m \dot n_i}{2\pi D} \left(\tilde{\mathcal{G}}(\mathbf 0) +\frac{1}{ \mathcal{F} R_i }\right)+ \sum_{j\neq i} \frac{m \dot n_j }{2\pi D}\mathcal{G}(\mathbf r_i-\mathbf r_j) = \rho_0-\rho_{\rm sat}.
\label{eq:matrixgrowthrates}
\end{equation}
This expression is valid for each droplet $i$, and thereby forms a closed set of linear equations for the $\dot n_i$.

\subsubsection{Global mass flux}
It is of interest to consider the global mass flux of vapor from the ceiling ($z=H$) to the bottom of the cell ($z=0$), which provides the total rate of condensed mass. Naturally, this flux must be equal to the total mass collected by all sinks, which per unit time per unit area reads $L^{-2} \sum_j m \dot n_j$. We can interpret this result using a horizontal Fourier decomposition of the diffusion problem, since only the zero-mode ({\em i.e.}, the average in the horizontal direction) contributes to the net flux from the ceiling across the cell towards the bottom. Indeed, the global flux as described by the zero-mode consists of a uniform vertical gradient from the average concentration $\rho_0$ at the top of the cell to the average at the substrate that we denote $\rho_{\rm eff}$. This flux $D(\rho_0 - \rho_{\rm eff})/H$ must be equal to $L^{-2} \sum_j m \dot n_j$, which readily gives the (average) substrate humidity effectively felt by droplets: 
\begin{equation}
\rho_{\rm eff} = \rho_0 - \sum_{j} \frac{ H m \dot n_j }{L^2 D}.
\label{eq:rhoeff}
\end{equation}
An alternative derivation of this result via a Fourier analysis of the Green's function (Appendix~\ref{app:electrostatics}) shows that the spatial average of $\mathcal{G}$ at $z=0$ is equal to $2\pi H/L^2$. By (\ref{eq:outer_asymptotics}), this gives the same expression for $\rho_{\rm eff}$. 

\section{Statistical model for breath figures}

\begin{figure*}[t]
\centering
\includegraphics{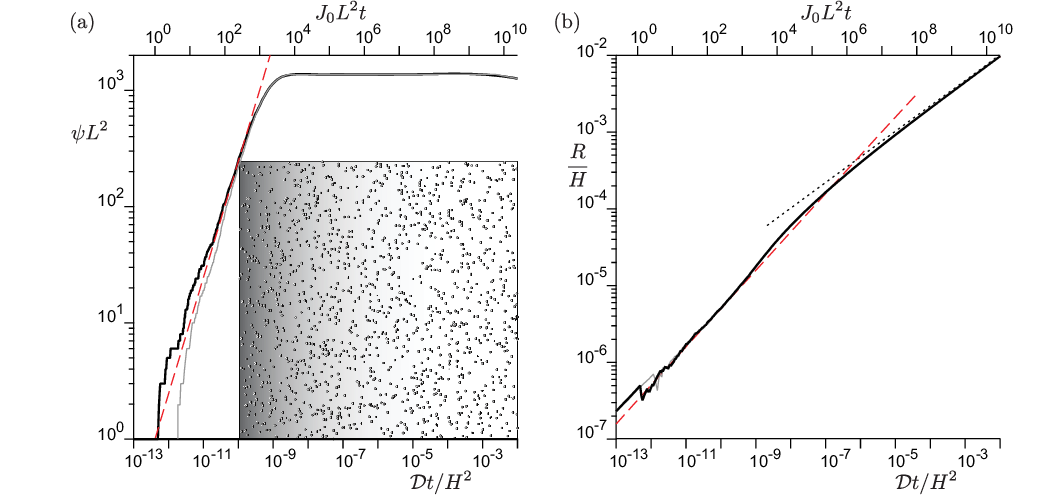}
\caption{Typical realisations of a stochastic simulation, without coalescence. 
(a) Number of drops $N=\psi L^2$ as a function of time $t$, rescaled by the diffusive time $H^2/\mathcal D$ (lower axis) or by the nucleation rate $J_0L^2$ (upper axis). The red dashed line corresponds to $\psi=J_0t$. Interaction between drops mediated by vapor diffusion leads to a small decrease of the number of the drops visible at long time. Inset: snapshot of a numerical realisation of a breath figure taken at $\mathcal D t/H^2=10^{-3}$, with the effective diffusion coefficient $\mathcal D$ defined in (\ref{eq:mathcalD}). (b) Root mean square drop radius $R = \sqrt{\langle R_i^2 \rangle}$, rescaled by $H$ plotted as a function of time $t$ (same rescaling as in panel (a)). The red dashed line corresponds to $R=\sqrt{\frac12( \mathcal R-1) \mathcal D t}$ (see (\ref{eq:MFRbisbis})} and the black dotted line to $R\propto t^{1/3}$. Numerical simulation is done at relative humidity $\mathcal R =\rho_0/\rho_{\rm sat} = 1.5$, the dimensionless energy gap $\chi=1.5$, horizontal box size $L=H$, $\mathcal F=1$ and $J_0 H^4/\mathcal D=2.4\;10^9$. The latter parameter is the ratio of the diffusion timescale to the (inverse) nucleation rate.
\label{fig:NaiveNumerics}
\end{figure*}

The above sections provide a many-droplet theory for breath figures (in the regime prior to coalescence). The theory includes the stochastic nucleation of new drops (Sec.~\ref{sec:SingleDropDescription}) and the deterministic diffusive growth of existing drops (Sec.~\ref{sec:MicroscopicDescription}), and consistently accounts for drop interactions via the vapor phase. We proceed by presenting results obtained from numerical simulations of the proposed model and show that breath figures evolve via two distinct regimes. Subsequently, we develop a mean field description that captures the numerical results without adjustable parameter and which is used for the interpretation of experiments in the companion paper \cite{PRL_companion}.

\subsection{Stochastic numerical simulations}

The numerical integration of the model is performed by two alternating steps. First, given the positions $\mathbf r_i$ and sizes $R_i$ of the drops, we compute the growth-rates $\dot n_i$ from (\ref{eq:matrixgrowthrates}); the drop radii $R_i$ are evolved accordingly. Second, the nucleation of a new drop is stochastically attempted using a Monte-Carlo step. In this step we randomly select a position on the substrate that is not yet covered by a droplet (i.e. $|\mathbf r-\mathbf r_i| > R_i$), and compute the local humidity according to \eqref{eq:outer_asymptotics}. From this, we accept/reject the nucleation in accordance to the statistical law \eqref{eq:rateenergyGap_final}. 

Figure~\ref{fig:NaiveNumerics} shows a typical result for the evolution of a single stochastic realisation of a breath figure. Model parameters are listed in the caption; the value of the relative humidity $\mathcal R=1.5$ is a typical experimental condition, $\mathcal F=1$ corresponds to hemispherical drops ($\theta=\pi/2$), while $\chi=1.5$ implies that the nucleation barrier is comparable to $k_BT$. The corresponding number of drops inside the cell $N=\psi L^2$ is plotted as a function of time in Fig.~\ref{fig:NaiveNumerics}(a). After starting with a single drop, it can be seen that the number of drops initially increases linearly in time, suggesting nucleation to occur at a constant rate, that is with a constant energy barrier. The dotted line corresponds to $\psi = J_0 t$, which gives a good description of the initial phase, where we introduced the initial nucleation rate
\begin{equation}\label{eq:defJ0}
J_0=\mathcal{J} \mathcal R^2 \exp \left(-\frac{\chi }{ \ln^2 \mathcal R}\right).
\end{equation}
The number of drops is observed to reach a saturation point around a well-defined average value, around $10^3$ in the present simulation. This indicates that nucleation is suppressed during the final stages of the process, revealing a reduction in supersaturation near the substrate. In fact, as will be quantified later, the effective humidity near the substrate, denoted as $\rho_{\rm eff}$, is found to decrease over the course of the condensation process. This decrease can be attributed to the increased volume of condensed liquid.

A similar crossover is observed for the evolution of the average drop size $R$, which we define as the root mean square drop radius $R = \sqrt{\langle R_i^2 \rangle}$. This average size is plotted as a function of time in Fig.~\ref{fig:NaiveNumerics}(b). Initially, the drops grow according to $R\sim t^{1/2}$, as is expected for isolated drops. The exact solution of the growth of a single isolated drop reads $R=\sqrt{(\mathcal R-1)\mathcal D t}$. However, the red dashed line corresponds to $R=\sqrt{\frac{1}{2}(\mathcal R-1)\mathcal D t}$, which provides a better fit: drops that nucleated at a later time $T$ have had less time to grow, which lowers the average radius of the ensemble. Given that the number of drops increases linearly in time, the average value of $\langle T\rangle = t/2$, so that, on average, the time to grow $t-\langle T\rangle = t/2$. This explains the factor 1/2 once multiple drops have nucleated. At the final stages of the simulation, the growth is sub-diffusive. Again, this is in line with a reduction of the supersaturation at the bottom of the cell, which slows down the growth just like it slowed down the nucleation.

\begin{figure*}[t]
\centering
\includegraphics{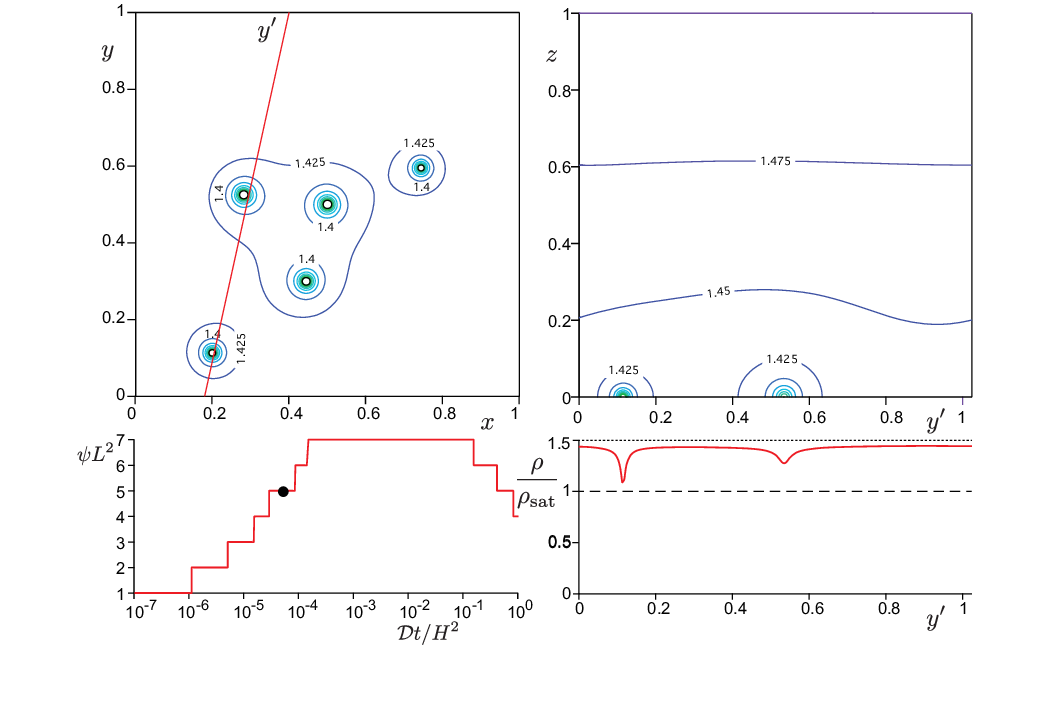}
\caption{
Humidity landscape in the first regime (low surface coverage): (a)~Top view of drops (black contours) and isocontours of water vapor density $\rho/\rho_{sat}$ at the substrate. (b)~Side view of isocontours $\rho/\rho_{sat}$ along the red line $y'$ drawn in panel (a). (c)~Number of drops $N=\psi L^2$ as a function of time $t$, rescaled by the diffusive time $H^2/\mathcal D$. The dot marks the time at which humidity fields in panels (a--b) are taken, $\mathcal D t/H^2=5\;10^{-5}$. (d)~Profile of water vapor density $\rho/\rho_{sat}$ at the substrate, plotted along the red line $y'$ drawn in panel (a). Numerical simulation is performed with an imposed relative humidity $\mathcal R =\rho_0/\rho_{\rm sat} = 1.5$, a dimensionless energy gap $\chi=1.5$, a horizontal box size $L=H$, $\mathcal F=1$ and $J_0 H^4/\mathcal D=1.8\;10^5$. The latter parameter is much lower than that in Fig.~\ref{fig:NaiveNumerics} to reduce the number of nucleated drops in the simulation box (other parameters remain unchanged).}
\label{fig:Fig9}
\end{figure*}
%Typical humidity field in the first regime (low surface coverage): (a) Drops (black contours) seen from the top and isocontours of the density of water vapor $\rho/\rho_{sat}$ at the substrate. (b) Isocontours of the density of water vapor $\rho/\rho_{sat}$ along the vertical plane that is indicated in red in panel a. (c) Number of drops $N=\psi L^2$ as a function of time $t$, rescaled by the diffusive time $H^2/\mathcal D$. The dot indicates the time at which the other panels are made. (d) Profile of the density of water vapor $\rho/\rho_{sat}$ at the substrate, along the red line shown in panel a. Numerical simulation is performed at relative humidity $\mathcal R =\rho_0/\rho_{\rm sat} = 1.5$, the dimensionless energy gap $\chi=1.5$, horizontal box size $L=H$, $\mathcal F=1$ and $J_0 H^4/\mathcal D=1.8\;10^5$. The latter parameter is much lower than that in Fig.~\ref{fig:NaiveNumerics} to nucleate less drops in the simulation box (other parameters are the same). Fields are taken at $\mathcal D t/H^2=5\;10^{-5}$.}

%
\begin{figure*}[t]
\centering
\includegraphics{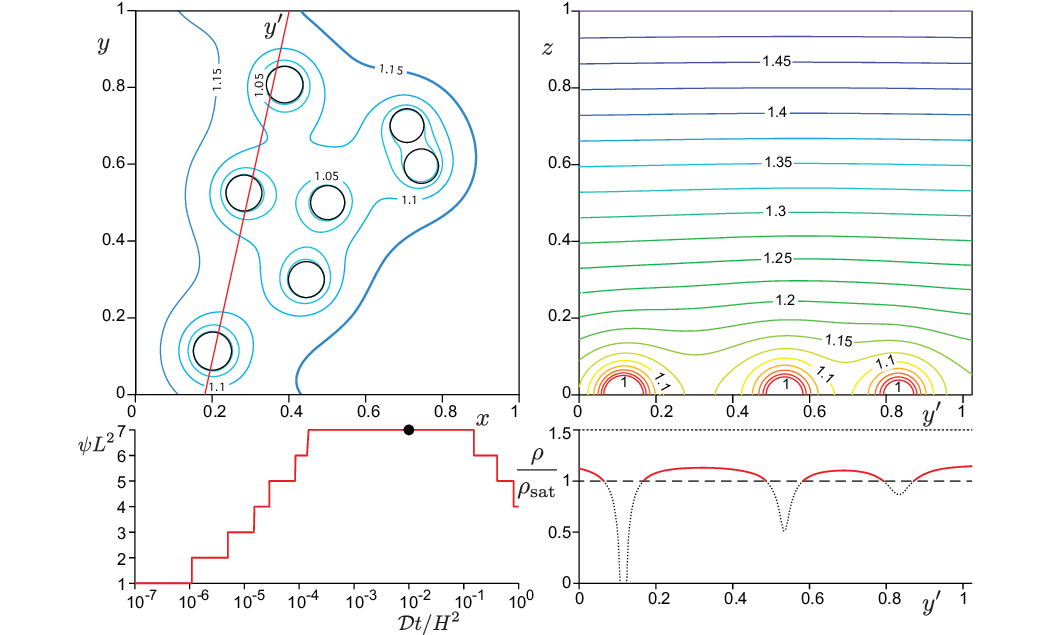}
\caption{Humidity landscape in the second regime (high surface coverage): (a)~Top view of drops (black contours) and isocontours of water vapor density $\rho/\rho_{sat}$ at the substrate. (b)~Side view of isocontours $\rho/\rho_{sat}$ along the red line $y'$ drawn in panel (a). (c)~Number of drops $N=\psi L^2$ as a function of time $t$, rescaled by the diffusive time $H^2/\mathcal D$. The dot marks the time at which humidity fields in panels (a--b) are taken, $\mathcal D t/H^2=10^{-2}$. (d)~Profile of water vapor density $\rho/\rho_{sat}$ at the substrate, plotted along the red line $y'$ drawn in panel (a). Parameters are the same as in Fig.~\ref{fig:Fig9}, but taken at a later time. }
\label{fig:Fig10}
\end{figure*}

%Typical humidity field in the second regime (high surface coverage): (a) Drops (black contours) seen from the top and isocontours of the density of water vapor $\rho/\rho_{sat}$ at the substrate. (b) Isocontours of the density of water vapor $\rho/\rho_{sat}$ along the vertical plane that is indicated in red in panel a. (c) Number of drops $N=\psi L^2$ as a function of time $t$, rescaled by the diffusive time $H^2/\mathcal D$. The dot indicates the time at which the other panels are made. (d) Profile of the density of water vapor $\rho/\rho_{sat}$ at the substrate, along the red line shown in panel a. Parameters are the same as in Fig.~\ref{fig:Fig9}, but taken at a later time, $\mathcal D t/H^2=10^{-2}$. }

The scenario that breath figures evolve according to two distinct regimes is very robust, and is seen in numerics for a broad range of parameter values. The underlying mechanism is illustrated in Figs.~\ref{fig:Fig9} and~\ref{fig:Fig10}, respectively showing humidity fields in the first and second regime. In comparison to Fig.~\ref{fig:NaiveNumerics}, the nucleation rate was reduced so that at most 7 drops appear in the simulation box to improve the readability of the humidity fields. During the first regime (illustrated in Fig.~\ref{fig:Fig9}), the average humidity at the substrate is close to the value imposed at the top of the cell ($\mathcal R=\rho_0/\rho_{\rm sat}=1.5$). Hence, the nucleation probability is large everywhere on the substrate, except in the very close vicinity of the drops. 
The situation is markedly different during the second regime (illustrated in Fig.~\ref{fig:Fig10}), during which the average humidity is close to $\rho_0/\rho_{\rm sat} =1$. In this regime, the humidity varies approximately linearly across the cell, and the drops effectively create a nearly-uniform saturated layer. Clearly, in this second regime, there no longer is a supersaturation at the substrate, and nucleation is suppressed. We note that the crossover to the second regime is seen to arise when the typical distance between drops $\lambda = \psi^{-1/2}$ is still much larger than the drop size $R$ (by about a factor 100 for the data in Fig.~\ref{fig:NaiveNumerics}). 
Thus, drops are small compared to their typical separation, as visually inferred from the inset in Fig.~\ref{fig:NaiveNumerics}(a), which imply that the second regime emerges well before coalescence starts to play a significant role.
%This fact that drops are small compared to their typical distance can also be inferred visually from the inset of Fig.~\ref{fig:NaiveNumerics}(a), and implies that the second regime emerges well before coalescence starts to play a significant role. }
In what follows, we develop a mean field model that quantitatively explain these features, including the crossover between the two regimes.

\subsection{Lattice of identical drops}

 The basis for the mean field description is an exact result that can be obtained for a regular lattice of identical drops. Treating all drops as equal, we write $R_i = R$, $\dot n_i=\dot n$, and $\rho_\infty^i = \rho_\infty$. The average concentration at the substrate, as defined by (\ref{eq:rhoeff}), then becomes

\begin{equation}\label{eq:lattice}
\rho_{\rm eff}= \rho_0 - \frac{H\psi m \dot n }{D}.
\end{equation}
We now assume that $\rho_\infty \approx \rho_{\rm eff}$, \textit{i.e.} the humidity experienced for each drop is approximated by the average humidity on the substrate. Numerical simulations show that the heterogeneities of the humidity field remain indeed very small with respect to the average, except in the very close vicinity of the drops (cf. Figs.~\ref{fig:Fig9} and~\ref{fig:Fig10}). With this, the growth equation (\ref{eq:GrowRateHydrodynamicRegime}) reads
\begin{equation}\label{eq:growthlattice}
m \dot n= 2 \pi \mathcal{F} R D\left(\rho_{\rm eff}-\rho_{\rm sat}\right).
\end{equation}
Inserting into (\ref{eq:lattice}), we find
\begin{equation}
 \rho_{\rm eff} - \rho_{\rm sat}= \frac{\rho_0-\rho_{\rm sat}}{1+2\pi\mathcal{F}\psi HR},
\label{eq:averagehumidity}
\end{equation}
which offers an explicit expression for the supersaturation at the surface. 

Equation (\ref{eq:averagehumidity}) reveals an essential feature for the evolution of breath figures: the supersaturation at the substrate, $\rho_{\rm eff} - \rho_{\rm sat}$, will decrease over time owing to the increase of droplet density $\psi$ and droplet size $R$. Specifically, we see the emergence of a dimensionless combination, 

\begin{equation}
\psi HR= \frac{HR}{\lambda^2},
\end{equation}
that quantifies the crossover from a regime of low surface coverage to a regime high surface coverage. In the low-density limit, $\psi HR \ll 1$, we find $\rho_{\rm eff} \approx \rho_0$ so that the supersaturation at the substrate is $\rho_0 - \rho_{\rm sat}$. This is the regime of non-interacting drops, in line with the early phase in Fig.~\ref{fig:NaiveNumerics}. Conversely, the high-density limit $\psi HR \gg1$ describes the second regime for which the bottom of the cell is no longer supersaturated: nucleation is arrested and droplet growth is subdiffusive. In line with stochastic numerics, the typical distance between drops at the moment of crossover $\lambda \sim \sqrt{HR}$, which is much larger than the typical drop size.

\subsection{Mean field model}

\subsubsection{Formulation}

We now formulate a mean field model for breath figures, consisting of evolution equations for the number density $\psi(t)$ and the average drop size $R(t)$. The latter is defined as $R = \sqrt{\langle R_i^2 \rangle}$. The nucleation rate (\ref{eq:rateenergyGap_final}) depends on the local humidity at the substrate. The mean field approximation consists of replacing the local humidity by the effective humidity $\rho_{\rm eff}$, so that the mean field equation for the number of drops takes the form,
\begin{equation}
 \frac{\mathrm{d}\psi}{\mathrm{d}t} = \mathcal{J} \left(\frac{ \rho_{\rm eff} } { \rho_{\mathrm{sat}}} \right)^2 \exp\left(-\frac{\Delta \Phi^*(\rho_{\rm eff} \big/ \rho_{\mathrm{sat}})}{k_B T}\right) 
 = {\mathcal J} \left(\frac{\rho_{\rm eff}}{\rho_{\rm sat}}\right)^2 \exp\left(- \frac{\chi}{\ln^2 \left( \rho_{\rm eff}/\rho_{\rm sat}\right)}\right)
\end{equation}
For $\rho_{\rm eff}$, we take the result obtained for the periodic lattice of identical drops (\ref{eq:averagehumidity}), which we rewrite as
\begin{equation}\label{eq:rhoeffmf}
 \rho_{\rm eff} = \frac{\rho_0+2\pi\mathcal{F}\psi HR\rho_{\rm sat}}{1+2\pi\mathcal{F}\psi HR}.
\end{equation}
Then, we assume the effective humidity (\ref{eq:rhoeffmf}) also determines the growth of individual drops. This amounts to setting $\rho_\infty^i = \rho_{\rm eff}$ in (\ref{eq:growthRsquared}), which fixes the growth of $R$ at constant $\psi$. However, the nucleation of new droplets also affects the average radius: the continuous addition of drops of vanishing volume at later times lead to a reduction of the average radius. We have seen (in the discussion around Fig.~\ref{fig:NaiveNumerics}) that during the initial phase where $\psi \sim t$, the average time at which a droplet nucleates is $\langle T \rangle = t/2$. Hence, the average radius will grow as $R^2 = \frac{1}{2}(\rho_{\rm eff}/\rho_0 - 1 ) \mathcal Dt$, where the factor $1/2$ accounts for the addition of freshly nucleated drops at later times. We capture the effect of a variable $\psi$ by proposing the mean field growth equation:
\begin{equation}
\frac{\mathrm{d} \left( \psi R^2\right)}{\mathrm{d}t} = \left( \frac{\rho_{\rm eff} }{ \rho_{\rm sat}}-1 \right) \mathcal D \psi.
\label{eq:growthmeanfield}
\end{equation}
This equation directly mirrors the result for isolated drops \eqref{eq:growthRsquared}, but contains additional factors $\psi$. Indeed, these extra factors are harmless at constant $\psi$, as is observed during the later stages. However, the incorporation of these factors, which are initially $\psi = J_0 t$, ensures the correct early-time result. 

\subsubsection{Dimensionless mean field model}

We now present the mean-field equations in their dimensionless forms. The model predicts the number of drops per unit area $\psi$, the average droplet radius $R$, the effective humidity at the substrate $\rho_{\rm eff}$, each of which is evolved over time. In the simulations, we used the size of the cell $H$ as the length-scale, and $H^2/\mathcal D$ as the time-scale. Aside from geometric parameters, the dimensionless parameters include the relative humidity $\mathcal R=\rho_0/\rho_{\rm sat}$, the energy gap $\chi$, and the ratio of timescales $J_0 H^4/\mathcal D$. The latter can be scaled out by choosing appropriate dimensionalization, using the fact that the crossover between the low-coverage and high-coverage limits occurs when $\mathcal{F}\psi RH$ is order unity. Specifically, we propose a rescaling where $\mathcal{F}\psi RH=\tilde \psi \tilde R$, where tildes denote dimensionless variables. Imposing that $R^2/t$ scales with $\mathcal D$, and $\psi/t$ with $J_0$, we obtain the following non-dimensionalization:
\begin{eqnarray}
 R & =&\left(\frac{\mathcal D}{J_0 \mathcal{F} H} \right)^{1/3} \, \tilde R, \label{eq:tildeR}\\
\psi & =& \left(\frac{J_0}{\mathcal D \mathcal{F}^2 H^2} \right)^{1/3} \ \tilde \psi , \label{eq:tildepsi} \\
 t & =& \left(\mathcal D J_0 ^2 \mathcal{F}^2 H^2 \right)^{-1/3} \, \tilde t , \label{eq:tildet} \\
 \rho_{\rm eff} & =& \rho_{\rm sat} \, \tilde{ \rho}_{\rm eff} , \label{eq:tilderho}
\end{eqnarray}
where tildes indicate dimensionless variables. The corresponding dimensionless mean field equations become:
\begin{eqnarray}
 \frac{\mathrm{d} \tilde \psi}{\mathrm{d}\tilde t}& =&\left( \frac{\tilde {\rho}_{\rm eff}}{\mathcal R} \right)^2 \exp \left(\frac{\chi }{ \ln^2 \mathcal R} -\frac{\chi }{ \ln^2 \tilde{\rho}_{\rm eff}} \right), \label{eq:MFpsi}\\
\frac{\mathrm{d} \left( \tilde {\psi} \tilde {R}^2\right)}{\mathrm{d} \tilde t} &=& \tilde {\psi} \left(\tilde{\rho}_{\rm eff}-1\right), \label{eq:MFR}\\
\tilde{\rho}_{\rm eff} &= &\frac{\mathcal R +2\pi \tilde{\psi} \tilde{R}}{1+2\pi \tilde{\psi} \tilde{R}} \label{eq:MFrho}.
\end{eqnarray}
The mean field model has thus been reduced to depend only on two dimensionless parameters: the dimensionless energy gap $\chi$, and the relative humidity imposed at the top of the cell $\mathcal R = \rho_0/\rho_{\rm sat}$.

\section{Results}

\subsection{Stochastic numerics versus mean field model}
\begin{figure*}[t]
\centering
\includegraphics{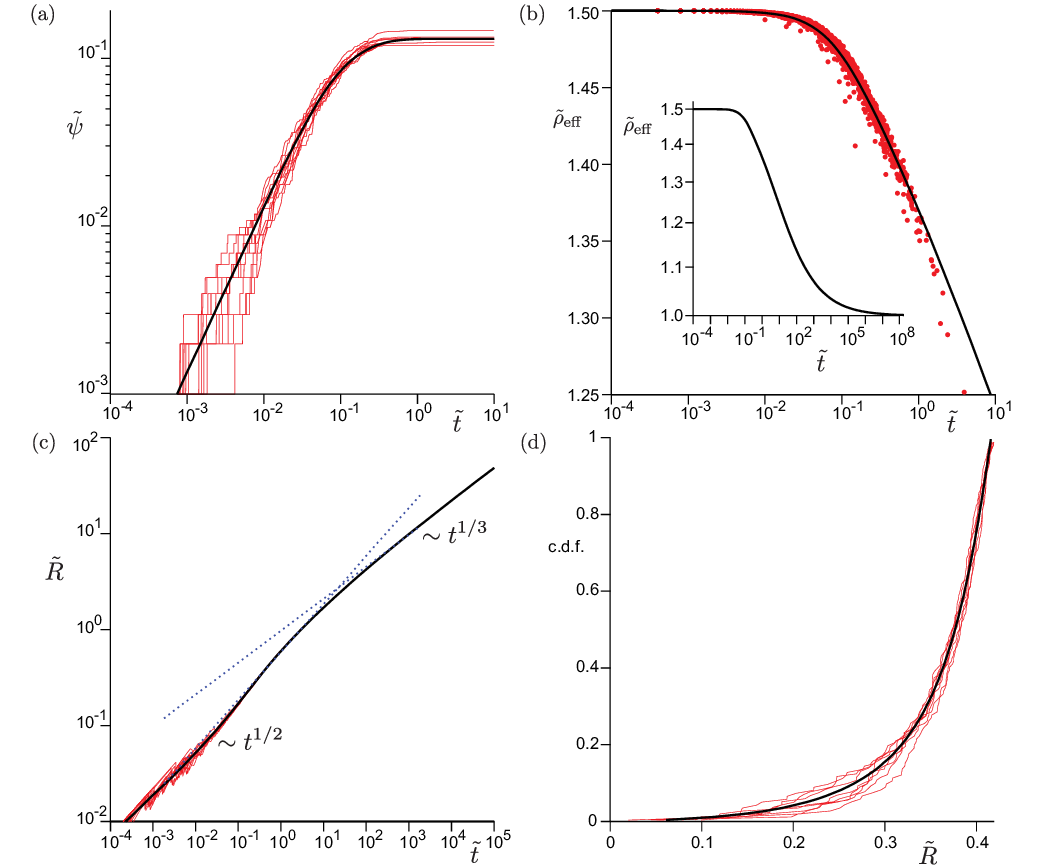}
\caption{Different realisations of the stochastic model (red) are compared to the mean field model (black) without adjustable parameters. The dimensionless parameters are set to $\mathcal R=1.5$ (a typical experimental condition) and $\chi = 1.5$ (which would correspond to water drops with $\theta=22^\circ$, that is the expected value on a PDMS melt in \cite{PRL_companion}). Variables carrying the $\tilde{(\,)}$ notation are nondimensionalized according to equations (\ref{eq:tildeR}-\ref{eq:tilderho})}. (a) Drop density $\tilde \psi$ as a function of time $\tilde t$. (b) Effective humidity at the substrate $\tilde \rho_{\rm eff}$ as a function of time $\tilde t$. Red dots correspond to samples at random locations at the substrate during the simulations, while the black line corresponds to the mean field prediction. The average over many realisations cannot be distinguished from the mean field prediction. Furthermore, the dispersion is observed to decrease with the drop density. The inset provides the mean field result over a larger time range, showing the transition from $\tilde \rho_{\rm eff}=\mathcal R=1.5$ to the fully saturated regime where $\tilde \rho_{\rm eff} \to 1$. (c) Root mean square drop size $\tilde R$ as a function of time $\tilde t$. Dotted lines indicate the analytical asymptotes of the mean field model (see text). (d) Cumulative distribution of drop sizes at $\tilde t=0.2$. For both panels, the average over many realisations cannot be distinguished from the mean field prediction. For this particular time, approximately 80\% of the drops fall within 10\% of the largest drop size.
\label{fig:StochasticModelpsi}
\end{figure*}

A comparison between different realizations of the stochastic simulations and the mean-field model (integrated numerically) is presented in figure~\ref{fig:StochasticModelpsi}, which shows the time evolution of the (dimensionless) droplet density $\tilde \psi$, the effective humidity at the substrate $\tilde \rho_{\rm eff}$, and the average radius $\tilde R$. In all cases, the average of many stochastic realizations of breath figures is indistinguishable from the mean-field prediction. The success of the mean-field model can be rationalized by examining Fig.~\ref{fig:StochasticModelpsi}(d), which displays the corresponding cumulative distribution function (c.d.f.) of drop sizes at a time within the saturated regime. The c.d.f. is directly obtained by ordering the $N$ drops by increasing radius and assigning the value $i/N$ to the $i^{\rm th}$ drop. For the mean-field model, a list of drops is gradually created in a deterministic way by creating a new drop when $\psi L^2$ crosses an integer value. The radius of all drops is increased according to the mean-field law:
\begin{equation}
\frac{\mathrm{d} \tilde {R}^2}{\mathrm{d} \tilde t} = \left(\tilde{\rho}_{\rm eff}-1\right), \label{eq:MFR2}
\end{equation}
The c.d.f. is calculated in the same way as for the simulations. The distributions reveal that the droplets become increasingly monodisperse as they grow at the same rate without any creation of new drops. This weak polydispersity aligns with experiments reported in the companion paper \cite{PRL_companion}, which validates our mean-field model to explain the monodisperse breath figure patterns obtained in diffusion-controlled condensation experiments on smooth substrates. In the remainder, we analytically derive asymptotic predictions from the mean-field model (\ref{eq:MFpsi},\ref{eq:MFR},\ref{eq:MFrho}).

\subsection{Regimes}
The dynamics of $\tilde \psi$ and $\tilde R$ exhibits two distinct regimes, the origin of which lies in the expression for effective humidity of the substrate (\ref{eq:MFrho}). As is illustrated in the inset of Fig.~\ref{fig:StochasticModelpsi}(b), the effective humidity $\tilde \rho_{\rm eff}$ starts at $\mathcal R$ (the relative humidity imposed at the top of the cell), but gradually decreases as the breath figures evolve. Indeed, the condensation of vapor locally reduces the average humidity at the substrate, until reaching a fully saturated regime where $\tilde \rho_{\rm eff} \to 1$. 
We first focus on the initial stages, the low-density limit, during which $\tilde \psi \tilde R \ll 1$. In this regime, (\ref{eq:MFrho}) reduces to $\tilde \rho_{\rm eff} \simeq \mathcal R$, and (\ref{eq:MFpsi},\ref{eq:MFR}) reduce to
\begin{eqnarray}
 \frac{\mathrm{d} \tilde \psi}{\mathrm{d}\tilde t}& \simeq & 1 , \label{eq:MFpsibis}\\
\frac{\mathrm{d} \tilde {\psi} \tilde {R}^2}{\mathrm{d} \tilde t} &\simeq& \tilde {\psi} \left(\mathcal R -1\right). \label{eq:MFRbis}
\end{eqnarray}
In this regime the nucleation occurs at a constant rate, which allows for an exact solution 
\begin{eqnarray}
 \tilde \psi& \simeq & t, \\
\tilde R^2 &\simeq& \frac{1}{2} \left( \mathcal R-1\right) t.\label{eq:MFRbisbis}
\end{eqnarray}
These results captures the early-time asymptotics of the breath figures, as seen in Fig.~\ref{fig:StochasticModelpsi}(a) and~\ref{fig:StochasticModelpsi}(c) and already indicated as dashed lines in Fig.~\ref{fig:NaiveNumerics}(a,b).\\
A second regime arises when $\tilde \psi \tilde R \gg 1$, for which $\tilde \rho_{\rm eff} \to 1$. In this limit, the right hand side of (\ref{eq:MFpsi}) vanishes so that no new drops are nucleated and $\tilde \psi$ reaches a plateau value that we indicate by $\tilde \psi_0$. Note, however, that the arrest of nucleation already happens long before the humidity reaches the fully saturated state. In Fig.~\ref{fig:StochasticModelpsi}(a), the plateau of $\tilde \psi$ is reached at $\tilde t \sim 10^{-1}$, when the effective humidity has only dropped by a small amount to $\tilde \rho_{\rm eff} \approx 1.45$ [cf. Fig.~\ref{fig:StochasticModelpsi}(b)]. Indeed, such a relatively small reduction of humidity is sufficient to dramatically slow down the nucleation process, owing to the Boltzmann factor that implies an exponential sensitivity of the nucleation rate. For the dynamics of the average radius $\tilde R$, the late-time asymptotic regime is reached significantly later, around $\tilde t \sim 10^3$ in Fig.~\ref{fig:StochasticModelpsi}(c), \textit{i.e.} once the humidity has dropped to values close to $\tilde \rho_{\rm sat} =1$. This regime corresponds to $\tilde \psi \tilde R \gg 1$, for which (\ref{eq:MFR}) reduces to:
\begin{eqnarray}
\frac{\mathrm{d}\tilde \psi \tilde {R}^2}{\mathrm{d} \tilde t} &\simeq& \frac{\mathcal R - 1}{2\pi \tilde R}. \label{eq:MFRtres}
\end{eqnarray}
Noting that $\tilde \psi = \tilde \psi_0$ is constant in this regime, this equation can be integrated to
\begin{equation}\label{eq:volumetric}
\frac{4\pi}{3}\tilde R^3 \simeq (\mathcal R - 1)\frac{\tilde t}{\tilde \psi_0}.
\end{equation}
This result implies a linear growth of the average droplet volume, corresponding to a sub-diffusive asymptotics $\tilde R \sim \tilde t^{1/3}$, and is indicated on Fig.~\ref{fig:StochasticModelpsi}(c). This growth law has been validated experimentally in the companion paper \cite{PRL_companion}. Given that the number of drops in the cell is constant in this regime, the linear growth of the volume can be interpreted from a constant flux of vapor condensing across the cell. This argument was already given in previous studies \cite{Viovy1988scaling,Rogers1988,Sokuler2010interspacing}. However, the present analysis explains why both the global flux and the number of drops remains constant: after a certain time, the humidity at the substrate is no longer supersaturated, which prevents nucleation and which (at the scale of the cell) gives rise to a uniform humidity gradient. Our analysis extends to situation where number of drops in the cell is not constant for instance owing to coalescence events or coarsening.

\subsection{Selection of the drop density}

\begin{figure*}[t]
\centering
\includegraphics{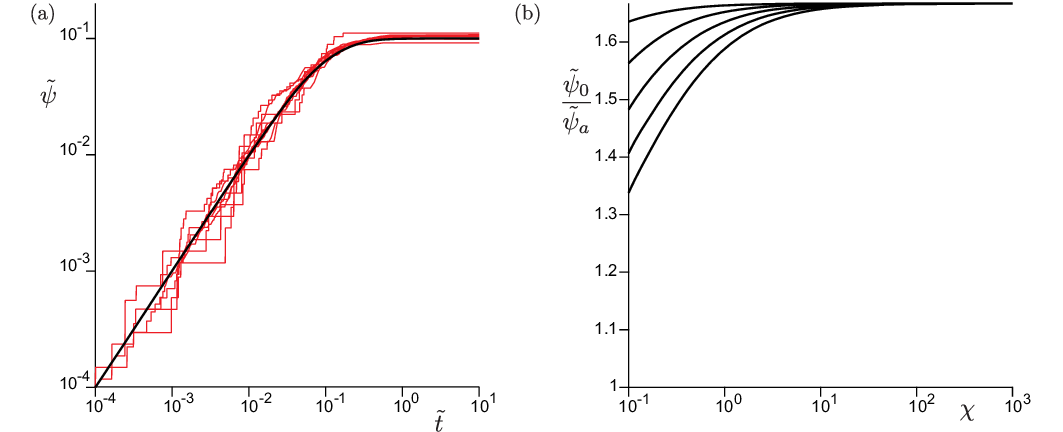}
\caption{(a) Different realisations of the stochastic model (red) compared to the mean field model (black) without adjustable parameters. Rescaling of the dimensionless drop density $\tilde \psi$ as a function of time for different $J_0 H^4/\mathcal D$ varied on a logarithmic scale from $2.4\;10^6$ to $2.4\;10^{12}$, for $L/H=1$, $\chi=1.5$ and $\mathcal R=1.5$. (b) Ratio of the selected drop density $\tilde \psi_0$ (plateau value in panel (a)) and the asymptotic prediction $\tilde \psi_a$ given by (\ref{eq:psia}), as a function of $\chi$. The chosen range of $\chi$ corresponds to water drops of increasing $\theta$ (roughly from $10^\circ$ to superhydrophic). Different curves correspond to different humidities $\mathcal R$.}
\label{fig:selection}
\end{figure*}

Finally, we wish to determine how the maximum number of drops $\tilde \psi_0$ depends on the two dimensionless parameters: the energy barrier $\chi$ and the imposed relative humidity $\mathcal R$. We have seen in Fig.~\ref{fig:StochasticModelpsi} that the value of $\tilde \psi_0$ is already reached at relatively high humidities, \textit{i.e.} when $\tilde \psi \tilde R \ll 1$. We accordingly expand the humidity and the nucleation equation as 
\begin{eqnarray}
\tilde{\rho}_{\rm eff} &\simeq &\mathcal R - 2\pi \tilde \psi \tilde R\left( \mathcal R - 1\right), \\
 \frac{\mathrm{d} \tilde \psi}{\mathrm{d}\tilde t}& \simeq & \exp \left(- \frac{4\pi \chi (\mathcal R-1)}{\mathcal R \ln^3 \mathcal R} \tilde \psi \tilde R\right). \label{eq:MFpsitres}
\end{eqnarray}
(\ref{eq:MFpsibis}) was derived assuming that the argument of the exponential factor was small, so that the nucleation rate is constant. For sufficiently large $\chi$, however, one can identify an asymptotic region that satisfies the hierarchy
\begin{equation}
\tilde \psi \tilde R \ll 1 \ll \frac{4\pi \chi (\mathcal R-1)}{\mathcal R \ln^3 \mathcal R} \tilde \psi \tilde R.
\end{equation}
In this regime the average drop size $\tilde R$ is still expected to grow diffusively, but the exponential factor is much reduced to slow down the nucleation. Given that the number of drops is approximately constant in this intermediate regime, the diffusive growth takes the form $\tilde R = \sqrt{(\mathcal R-1) \tilde t}$ (\textit{i.e.} without the factor $1/2$). Plugging this intermediate asymptotics in (\ref{eq:MFpsitres}), the integral can be explicitly carried out and gives:
\begin{equation}
\tilde \psi(\tilde t) \simeq \tilde \psi_0 - \frac{2}{c^2} \left(1+c\sqrt{\tilde t}\right) \exp \left(- c \sqrt{\tilde t}\right), \quad \mathrm{with} \quad c =\frac{4 \pi \chi (\mathcal R-1)^{3/2}}{\mathcal R \ln^3 \mathcal R} \tilde \psi_0 ,
\end{equation}
which reveals how $\tilde \psi_0$ is approached. When $\chi$ is assumed large, one indeed finds an exponentially fast approach to $\tilde \psi_0$, much before the product $\tilde \psi \tilde R$ has reached order unity. In this scenario, nucleation is effectively arrested once the exponential factor has decreased by some fixed factor. Hence, we can estimate the arrest time $\tilde \tau_a$ from the condition $c \sqrt{\tilde \tau_a} \sim 1$, which also leads to the estimation of the number density at arrest as $\tilde \psi_0 \sim \tilde \tau_a$. From this, we define the typical arrest density 
\begin{equation}\label{eq:psia}
\tilde \psi_a \equiv \frac{\mathcal R^{2/3}\;\ln^2 \mathcal R}{ \chi^{2/3} \left(\mathcal R - 1 \right)}, 
%\quad \mathrm{and} \quad
%\tilde \psi_a \sim \mathcal R^2 \tilde \tau_a \sim \frac{\mathcal R^{4/3} \ln^2 \mathcal R}{\chi^{2/3} \left(\mathcal R - 1 \right)}.
\end{equation}
which is therefore expected to capture the scaling law with respect to the model parameters $\chi$ and $\mathcal R$. We remark that this result implies
%$\tilde \psi_a \tilde R_a \sim 1/\chi$
\begin{equation}
\tilde \psi_a \tilde R_a\sim \frac{\mathcal R\;\ln^3 \mathcal R}{ \chi \left(\mathcal R - 1 \right),} 
\end{equation}
so that, indeed, a large energy barrier leads to the arrest of nucleation well before the effective humidity has dropped to the saturation value.

These relations are confirmed in Fig.~\ref{fig:selection}. In panel a, we first verify that the stochastic simulations for different $J_0H^4/\mathcal D$ (but fixed values $\chi$ and $\mathcal R$) collapse when introducing the dimensionless droplet density $\tilde \psi$ and dimensionless time $\tilde t$. Then, Fig.~\ref{fig:selection}(b) reports the arrest density $\tilde \psi_0$ for different values of $\chi$. The graph confirms that $\tilde \psi_0$ follows the asymptotic prediction $\tilde \psi_a$ given by (\ref{eq:psia}), for sufficiently large $\chi$. 

\section{Coarsening by coalescence}\label{sec:coarsening}
\label{sec:coalescence}

\subsection{Stochastic simulations including merging}
So far, we have considered the non-interactive and vapor-mediated interactive regimes assuming that the typical distance between drops $\lambda = \psi^{-1/2}$ remains much larger than the drop size $R$, that is before the occurrence of coalescence events. This is consistent for the description of the nucleation phase and its arrest. Namely, the stage where the number of drops is constant and drop sizes grow according to $R \sim t^{1/3}$ is reached when $\psi R H \lesssim 1$, so that the typical distance between drops $\lambda \sim \sqrt{RH} \gg R$. Hence, the probability of droplets to touch and merge is negligible during the phases described above. However, drops will continue to grow so that ultimately they will touch. To investigate this effect, we extend our numerical simulations by including coalescence, by assuming an instantaneous merging when two droplet radii overlap. When an overlap occurs between droplet $i$ and $j$, we instantaneously create a new droplet of radius $(R_i^{3}+R_j^{3})^{1/3}$ (ensuring volume conservation). The new droplet is placed at the centre of mass position of the two preceding drops (ensuring momentum conservation). When implementing the model in a code, only pairs of drops can merge. However, this can induce in cascade other mergings — a very rare situation in practice. The model of the vapor is still based on quasi-steady diffusion, i.e. assuming fast transients, so that within the simulations the vapor field instantaneously adapts after the merging. 

\begin{figure*}[t]
\centering
\includegraphics{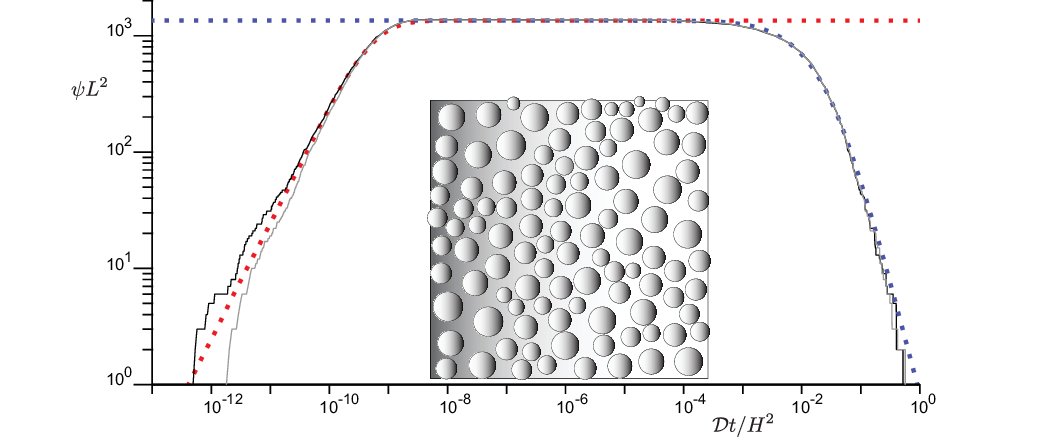}
\caption{Effect of coalescence on the number of drops $N=\psi L^2$ versus dimensionless time $\mathcal Dt/H^2$. Black and grey solid line: Stochastic numerical simulations for the same parameters as Fig.~\ref{fig:NaiveNumerics} and the same seed in the random number generator, but now including drop coalescence. Red dotted line: Prediction of the mean field model. Blue dotted line: Best fit by the coarsening theory (\ref{eq:psieq}). The inset shows the drop configuration during the coarsening phase at $N=100$.}
\label{fig:Coarsening}
\end{figure*}

Figure~\ref{fig:Coarsening} shows a typical evolution for the number of drops $\psi(t)$, in a simulation that includes nucleation, growth, and coalscence. As in Fig.~\ref{fig:NaiveNumerics}(a), we observe a fast nucleation stage, followed by a plateau that marks the arrest of nucleation. Then, drops continue to grow until they touch and merge, after which the number of drops start to decay, following approximately the scaling law $\psi \sim 1/t^2$. 

\subsection{Mean field model for self-similar coarsening}
\begin{figure*}[t]
\centering
\includegraphics{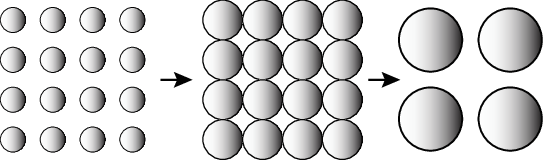}
\caption{Coarsening on a square lattice. Drops are assumed to grow according to (\ref{eq:defU}) until they touch. In this model, 4 drops are assumed to merge instantaneously while preserving volume, leading to a new square lattice. The product $\psi R^2=1/4$ just before coalescence while $\psi R^2=A=2^{-8/3}$ just after coalescence.}
\label{fig:SchematicCoarsening}
\end{figure*}

The observed scaling law for coarsening can be explained by assuming self-similar coarsening dynamics \cite{Fritter1988,Derrida1991,Meakin1992}. To illustrate the concept we first consider the idealised situation of a square lattice of identical drops. When in contact, drops are assumed to merge as is sketched in Fig.~\ref{fig:SchematicCoarsening}. The result is a new square lattice with a lattice unit that is increased by a factor 2. This operation instantaneously reduces the number of drops by a factor 4, while volume conservation implies that the drop radius after merging is increased by a factor $4^{1/3}$. Denoting $\psi_n$ as the number of drops per unit area after the $n$-th merging event, we thus have $\psi_n = \psi_0/4^n$. If we furthermore define $R_n$ as the droplet size immediately after coalescence, one verifies that for a square lattice $\psi_n R_n^2=A$ is constant, with $A=2^{-8/3}$.

To transform this coarsening into a temporal dynamics for $\psi(t)$, we need to determine the time in between two merging events. For this, we make use of the growth law in the saturated regime, given by (\ref{eq:volumetric}), which we write in dimensional form, 
\begin{equation}\label{eq:defU}
\psi \frac{d R^3}{dt} = \mathcal U, \quad \mathrm{with} \quad \mathcal U= \frac{3 \left(\rho_0 - \rho_{\rm sat} \right) \mathcal D}{4 \pi \mathcal F \rho_{\rm sat} H}.
\end{equation}
Here we introduced the characteristic velocity $\mathcal U$ that represents the volume flux per unit area that dictates the growth. We wish to compute the time interval between the merging events $n$ and $n+1$. In between the coalescence times $t_{n}$ and $t_{n+1}$, the droplet size grows from $R_n$ to $2^{1/3}R_n$. Using that $\psi_n$ is constant in between these events, we obtain from (\ref{eq:defU})
\begin{equation}\label{eq:tn}
t_{n+1} - t_n = \frac{\psi_n R_n^3}{\mathcal U}.
\end{equation}
Using $\psi_n R_n^2 = A$, we can thus express the time interval as
\begin{equation}\label{eq:tn}
t_{n+1} - t_n = \frac{A^{3/2}}{\psi_n^{1/2} \mathcal U} = 2^{n} \frac{A^{3/2}}{\psi_0^{1/2} \mathcal U} .
\end{equation}
which can be summed to
\begin{equation}
\left(\frac{\psi_0^{1/2} \mathcal U}{A^{3/2}} \right) \left(t_n - t_0 \right) = \sum_{k=0}^{n-1} 2^n = 2^n - 1.
\end{equation}
From this we infer the temporal evolution of $\psi_n$, as
\begin{equation}\label{eq:psiA}
\psi_n = \frac{\psi_0}{4^n} = \frac{\psi_0}{\left(1+ \frac{\psi_0^{1/2} \mathcal U}{A^{3/2}} (t_n-t_0) \right)^2}.
\end{equation}
Hence, the lattice model exhibits the sought-after scaling $\psi \sim 1/t^2$. In hindsight, this result could have been inferred on dimensional grounds. A merging event transforms the square lattice into a square lattice of larger size, which renders the problem self-similar. The resulting dynamics must display scale-invariance: the model indeed does not exhibit an intrinsic length scale, nor an intrisic timescale. The coupling of space and time is established via the growth law (\ref{eq:defU}), which involves a parameter $\mathcal U$ that has the dimensions of a velocity. Hence, the only way to construct a growing lattice size $\psi^{-1/2}$ is via the scaling $\sim \mathcal Ut$. Changing the type of lattice (e.g. from square to hexagonal) would only affect the value of $A$, not the scaling law.

Obviously, the lattice model is not intended as a realistic description of the stochastic numerics: Drops are not equally sized, not equally spaced, and merge in pairs of two rather than in pairs of four (cf. inset Fig.~\ref{fig:Coarsening}). If, however, we assume the coarsening patterns to be statistically self-similar, then the growth law (\ref{eq:defU}) again offers $\mathcal U$ as the sole dimensional parameter in the problem. For the same reason as for the lattice, this dictates the growth of the pattern length-scale to satisfy $\psi^{-1/2} \sim \mathcal Ut$. We thus anticipate (\ref{eq:psiA}) to be valid, but with a non-universal factor $A$. We therefore propose the form 
\begin{equation}
\psi=\frac{\psi_0}{(1+(t-t_0)/\tau_c)^2},
\label{eq:psieq}
\end{equation}
where $\tau_c$ should be of the order $\psi_0^{-1/2} \mathcal U^{-1}$. The fit of (\ref{eq:psieq}) is superimposed as the blue dashed curve in Fig.~\ref{fig:Coarsening}. The proposed form also provides an excellent fit of the experimental data obtained on soft gels, presented in the companion paper \cite{PRL_companion}. 

As a final comment, we note that in the stochastic simulations we do not encounter any renewed nucleation during the coalescence phase, in line with experiments on brushes \cite{PRL_companion}. We can indeed verify that the dimensionless surface coverage $\psi R H$ stays sufficiently large, so that nucleation remains inhibited. Namely, during the coalescence phase the product  $\psi R^2 = A$ remains constant, so that $\psi R H \sim  H/R$, which indeed is large during coarsening. This scenario becomes manifestly different in the presence of strong convection, for which $H$ is to be replaced by a thin diffusive boundary layer, as was qualitatively suggested in \citep{beysens2022physics}. Indeed, in many experiments reviewed in  \citep{beysens2022physics} there is continuous re-nucleation during the coalescence phase, in contrast to our diffusion-controlled experiments  \cite{PRL_companion}.

\section{Discussion}
This study presents a theoretical framework to clarify the formation of nearly monodisperse breath figures on defect-free substrates under a diffusion-controlled regime, as observed in the experimental companion paper \cite{PRL_companion}. Specifically, we address the question of how condensing vapor is distributed between the growth of existing droplets and the nucleation of new ones. We demonstrate that, following an initial rapid nucleation stage, dew droplets begin to compete for vapor absorption, leading to the establishment of a vapor-depleted boundary layer near the substrate. These collective effects, mediated by vapor diffusion, lead to two distinct regimes: a low-density nucleation regime ($\psi RH< 1$) and a high-density saturation regime ($\psi RH>1$) where droplet interactions commence. In the low-density nucleation regime, the effective humidity at the substrate remains close to the imposed relative humidity at the top of the cell, resulting in diffusive droplet growth and a consistent nucleation rate. However, as droplets grow and attain higher densities, the system shifts to the high-density regime. In this regime, the effective humidity experienced by droplets decreases, impeding their growth and ultimately halting nucleation due to the exponential sensitivity of nucleation rates to humidity variations. This transition signifies the beginning of the constant-drop-number regime, preceding the coalescence regime. As such, the model explains the experimental measurements (detecting droplets once they reached the micron size), which indeed exhibit a constant number of drops (until coalesce kicks in) \cite{PRL_companion}. 

Our mean-field model for the formation and evolution of breath figures yields predictions for the effective humidity field $\rho_{\rm eff}$, the average droplet radius $R$, and the droplet density $\psi$. Comparison of our theory with stochastic simulations reveals excellent agreement, particularly in the nucleation and growth phases, highlighting the robustness of the mean-field approximation. Specifically, our simulations have corroborated several key findings: the plateau in droplet number $\psi\rightarrow\psi_0$, the sub-diffusive growth law $R \sim t^{1/3}$ or linear growth of droplet volume $\psi R^3 \sim t$ in the interactive regime $\psi RH>1$, and the nearly monodisperse distribution of droplet sizes, characterized by a cumulative distribution function (c.d.f.) skewed towards larger droplets. Both our theory and stochastic simulations are consistent with the dynamics observed in experimental findings \cite{PRL_companion}. 

The mean-field model offers insights into the maximum droplet density, $\psi_0$, which is determined by dimensionless parameters: the energy barrier $\chi$ and the imposed relative humidity $\mathcal{R}$. Our analysis indicates that $\psi_0$ is attained before the product $\psi R H$ reaches $1$, primarily due to the exponential suppression of nucleation rates. The derived scaling law for $\psi_a$ (\ref{eq:psia}) is confirmed by stochastic simulations, demonstrating the predictive capability of our model. 
This result is of particular importance for the quantitative interpretation of the experiments in the companion paper \cite{PRL_companion}, which involves various types of deformable substrates. The deformability allows for drops to partially ``sink" into the substrate, changing the ratio of volume to surface coverage, an effect that can be captured in the presented theory by a modification of $\mathcal F$. The real challenge, however, is that the nucleation rates and the associated energy barriers are not known a priori for soft substrates; the model predictions presented here can be used back-track these microscopic parameters from the measured $\psi_0$.

We have also integrated the late-stage evolution of droplet numbers into our model. By incorporating coalescence events in our simulations, we identified a coarsening phase where droplet numbers decrease over time as they merge. This process follows a scaling law $\psi \sim 1/t^2$ indicative of self-similar coarsening dynamics. The mean-field model for self-similar coarsening, based on the growth law and assuming statistical self-similarity, accurately described the evolution of droplet density during this phase. The proposed scaling for $\psi (t)$ (\ref{eq:psieq}) provided an excellent fit to both the simulation and experimental data \cite{PRL_companion}.

The monodisperse breath figure patterns and growth laws predicted by our theory are corroborated by experimental observations detailed in our companion paper \cite{PRL_companion}, and by the observed patterns obtained on smooth substrates, liquid \cite{Zhang2020,Steyer1990,Steyer1993,Nepomnyashchy2006,Guha2017}, liquid-infused surfaces \cite{Anand2015,Sharma2022} or polymeric substrates \cite{Briscoe1991,Ge2020,Sokuler2010}, further validating our approach. However, our model relies on the formation of a diffusive vapor-depleted boundary layer near the substrate, which may encounter challenges in configurations where this boundary layer cannot form or is unpredictable. Depending on the substrate orientation and the humidity injection system (its flow rate), the flow can exhibit convective or turbulent characteristics, leading to heterogeneous mixing \cite{Villermaux1999,Villermaux2019}. To maintain stable stratification and prevent natural convection, the substrate must be cooled from below, as cold vapor is denser than warm vapor in the vicinity of the substrate. This condition is always fulfilled in experiments conducted on liquid layers to prevent liquid drainage or destabilization. However, it excludes studies where condensation experiments are conducted on vertical substrates for water harvesting or on downward-facing horizontal substrates. These factors, along with the possible presence of surface defects and the eventual occurrence of coalescence, may explain why in some experimental studies, conducted on smooth \cite{family1989kinetics,Sikarwar2011}, polymeric films or gels \cite{Katselas2022,Rose2002,Phadnis2017,Sokuler2010,Stricker2022universality, Kolb1989,Leach2006,Trosseille2019} and liquid-infused surfaces \cite{Ge2020,Anand2015,Lavielle2023}, the dew patterns exhibit polydispersity.\\

In conclusion, our model provides valuable insights into the mechanisms underlying the formation of quasi-monodisperse dew patterns on defect-free surfaces. These findings have important implications for controlling dew pattern formation in various applications, such as designing nano-emulsions with innovative optical properties or patterning substrates for microfabrication. By understanding the fundamental physics of breath figure formation, we can develop new strategies to harness condensation for a wide range of technological applications.\\

\noindent {\bf Acknowledgments. }The authors gratefully acknowledge discussions with U. Thiele and C. Henkel. Final support from NWO Vici (No. 680-47-632), and DFG (Nos. SN145/1-1 within SPP 2171). 

\section*{Appendix}
\begin{appendix}
\new{\section{Energy barrier}\label{app:ExternalG}
\subsection{Thermodynamic potential}
We recall here the main steps of the derivation of the thermodynamic potential relevant to nucleation in an oversaturated atmosphere, as can be found e.g. in \cite{Frenkel,diu2007thermodynamique}, and compute the resulting energy barrier. We consider a closed system that consists of a liquid drop and the surrounding vapor, as illustrated in Fig,~\ref{fig:appendixA}. It is connected to a reservoir with fixed pressure $p_0$ and temperature $T_0$ (thermostat), allowing for exchange of volume and of heat; the number of molecules $N$ in the system  remains constant. We therefore build the Gibbs external free energy $\Phi$ defined by:
$$
\Phi = U  - T_0 S + p_0 V.
$$
where $U$, $S$, and $V$ are the overall internal energy, entropy and volume of both liquid and vapor phases. During an elementary transformation, the system exchanges elementary work $\delta W$ and heat $\delta Q$ with the reservoir. The reservoir being a thermostat at controlled pressure, its transformation is reversible even when that of the system is not. The work exchanged by the reservoir reads $-\delta W=p_0 dV$. The heat exchanged by the thermostat reads $-\delta Q=T_0 dS_{\rm th}$ where $S_{\rm th}$ is the entropy of the reservoir. The variation of the total internal energy of the closed system therefore reads $dU=\delta W+\delta Q=-p_0 dV-T_0 dS_{\rm th}$. The total entropy of the system plus reservoir reads $S_{\rm tot}=S_{\rm th}+S$, and one thus verifies that $d\Phi=-T_0 dS_{\rm tot}$. As the total entropy $S_{\rm tot}$ is maximal at equilibrium, the external free energy $\Phi$ is minimal, which is therefore a correct thermodynamic potential for this problem.
 \begin{figure}
\centering
\includegraphics[width=.5\textwidth]{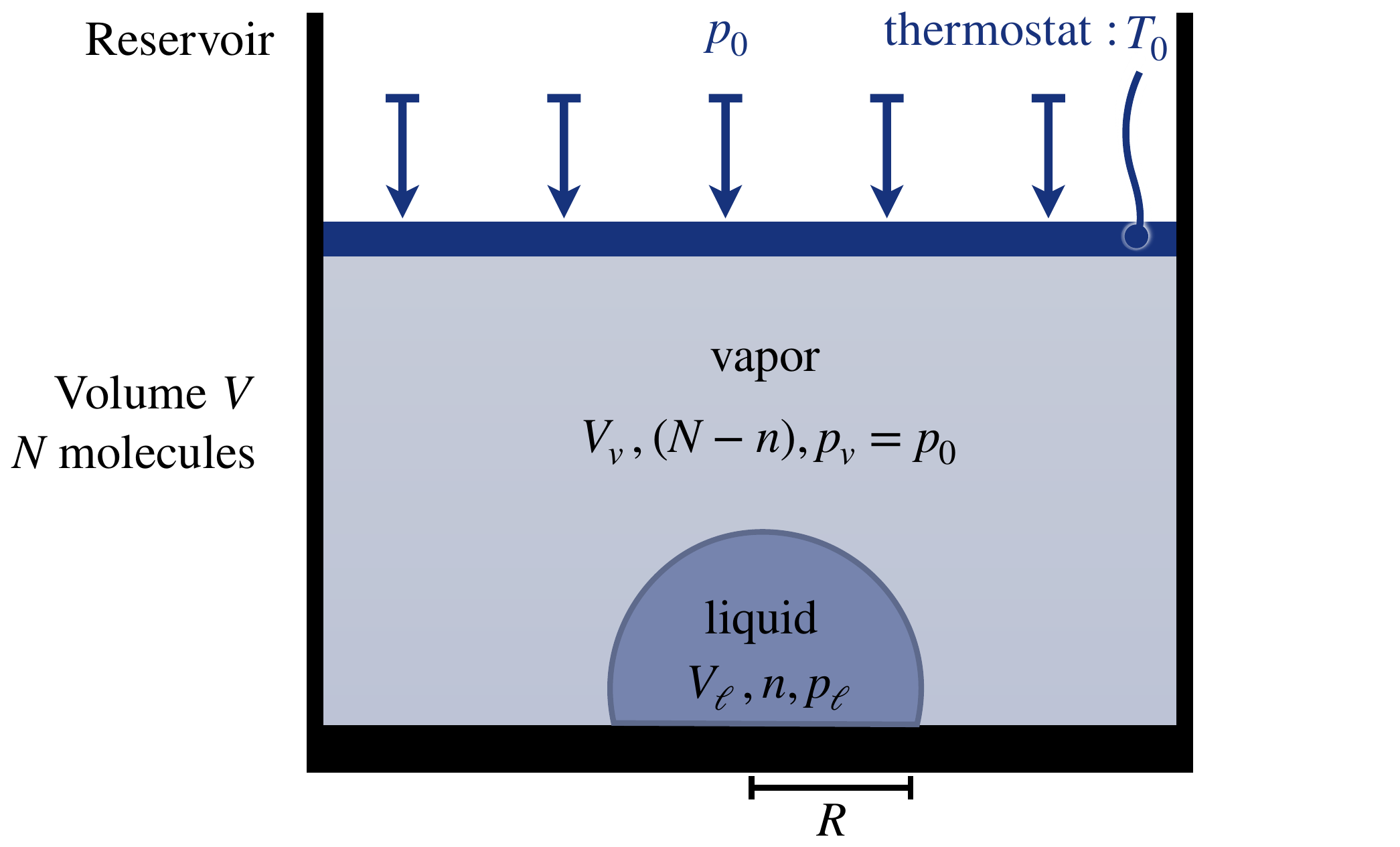}
\caption{\new{Thermodynamic system considered. A cell of volume $V$ and $N$ molecules, is divided into a vapor and a liquid phase. The system is in constant with a reservoir at pressure $p_0$, so that the volume $V$ is allowed to change; the number of molecules $N$ is constant. The vapor phase has a volume $V_v=V-V_\ell$, $(N-n)$ molecules and its pressure $p_v=p_0$ at equilibrium. The liquid phase has a volume $V_\ell$, $n$ molecules and a pressure $p_\ell$.  The system is assumed to be at constant temperature $T_0$ of the thermostat.}}
\label{fig:appendixA}
\end{figure}

\subsection{Kelvin effect}
Having established the relevant potential $\Phi$, let us first show that the Laplace pressure and the Kelvin effect are recovered at equilibrium. The temperature $T_0$ and the total number of molecules $N$ in the system are kept constant. The degrees of freedom of the system are the number of molecules inside the drop $n$, the volume of the drop $V_\ell$, the volume of the vapor $V_v= V-V_\ell$. The number of molecules in the vapor $N-n$ is not independent, since $N$ is fixed. With this, we differentiate $\Phi$ as:
\begin{eqnarray}\label{eq:differential}
d\Phi = d(F + p_0 V) &=&   (\mu_\ell - \mu_v)dn  - p_\ell dV_\ell - p_0 dV_v + dF_s + p_0 dV    \nonumber \\
&=&(\mu_\ell - \mu_v)dn + (p_0-p_\ell) dV_\ell +\gamma dA.
\end{eqnarray}
Here we introduced $F_s$ as the interfacial free energy, which for an isolated drop $F_s  = \gamma A$ where $A$ is the surface area of the drop. The change $dA$ is not independent of $dV$. For a spherical droplet (assuming homogeneous nucleation to simplify geometrical prefactors), we have $dA = \kappa dV_\ell$, where $\kappa$ is twice the mean curvature of the sphere. Hence, the differential (\ref{eq:differential}) leads to two equilibrium conditions, respectively, associated to $dn$ and $dV$:
\begin{equation}\label{eq:KelvinLaplace}
\mu_\ell(T_0,p_\ell) = \mu_v(T_0,p_0), \quad \mathrm{and} \quad p_\ell - p_0 = \gamma \kappa.
\end{equation}
At equilibrium, we thus recover the pressure difference by Laplace's law and equality of chemical potentials. Given the difference in pressure, we evaluate the liquid chemical potential as 
$\mu_\ell(T_0,p_\ell) =  \mu_\ell (T_0,p_0) + v_\ell (p_\ell - p_0)$, where $v_\ell= \frac{\partial \mu_\ell}{\partial p}$ is the volume per liquid molecule. For an incompressible fluid, $v_\ell$ is independent of pressure and the expression is exact (i.e. there are no higher order terms in the expansion). Taking the saturation pressure $p_{\rm sat}$ as a reference (like we will do for the vapor), we can equivalently write
$\mu_\ell(T_0,p_\ell) = \mu_\ell (T_0,p_0) + v_\ell (p_\ell - p_0) + v_\ell(p_0 - p_{\rm sat})$. The shift in chemical potential with respect to a planar liquid-vapor interface at temperature $T_0$ and pressure $p_0$ must be matched by the same shift in the vapor phase. Treating the vapor as an ideal gas, with saturation pressure above a flat surface $p_{\rm sat}$, one finds: $\mu_v(T_0,p_0) = \mu_v(T_0,p_{\rm sat}) +  (k_B T/v_\ell) \ln (p_0/p_{\rm sat})$. Equality of chemical potentials $\mu_\ell(T_0,p_\ell) = \mu_v(T_0,p_0)$ then gives the selection of the droplet's curvature at equilibrium:
\begin{equation}
\gamma \kappa = \frac{k_B T}{v_\ell} \ln (p_0/p_{\rm sat}) + p_{\rm sat} - p_0,
\end{equation}
which is the Kelvin effect, or the Ostwald-Freundlich equation. The calculation can be adapted to spherical caps in contact with a substrate by adding the relevant surface energies, which alters the geometric factors; besides the above equilibrium conditions, one will in addition find Young's law for the contact angle.}

\new{\subsection{Energy barrier}
We wish to calculate the energy barrier for nucleation, $\Delta \Phi^*$, which is the maximum of $\Delta \Phi$ measured with respect to the state where the whole fluid is a homogeneous vapor state at  temperature $T_0$ and pressure $p_0$. As the temperature is homogeneous $T=T_0$ in both phases at equilibrium, we can express $\Delta \Phi$ as a function of the free energies $F_\ell$ and $F_v$ of $n$ molecules in the liquid and vapor phases, respectively. To create a drop, $n$ molecules have passed from a vapor state at temperature $T_0$ and pressure $p_0$ to a liquid state  at temperature $T_0$ and pressure $p_\ell$. They were occupying a volume $V_v$ and had a free energy $F_v$ in the vapor phase, which is replaced by a volume $V_\ell$ with a free energy $F_\ell$ in the liquid state. Combined, we get a change $\Delta \Phi$ that reads:
$$
\Delta \Phi=F_\ell\left(T_0, V_\ell, n\right) - F_v\left(T_0, V_v, n\right)  +p_0 \left( V_\ell(T_0,p_\ell,n) - V_v(T_0,p_0,n) \right)+F_s.
$$
In this expression we can recognize the Gibbs free energy of $n$ molecules of vapor, namely $n \mu_v\left(T_0, p_0\right) = F_v\left(T_0, V_v, n\right)+ p_0 V_v\left(T_0, p_0,n\right)$. It is instructive to also introduce the Gibbs free energy of the liquid droplet $n \mu_\ell(T_0,p_\ell) = F_\ell\left(T_0, p_\ell, n\right)+ p_\ell V_\ell \left(T_0, p_\ell \right)= F_\ell\left(T_0, p_\ell, n\right)+ n p_\ell v_\ell \left(T_0 \right)$, where we assumed the liquid to be incompressible. Then, we add and substract the term $n p_\ell v_\ell\left(T_0\right)$ to the expression for $\Delta \Phi$, so that
\begin{equation}\label{eq:Phi}
\Delta \Phi=n \left[ \mu_\ell\left(T_0, p_\ell\right)- \mu_v\left(T_0, p_0\right) \right] + n v_\ell(T_0) \left(p_0-p_\ell\right) +F_s.
\end{equation}
We then further express the chemical potential of the liquid, using $v_\ell$ is constant, which yields
$$
\begin{aligned}
\mu_\ell\left(T_0, p_0\right) 
%& = \mu_\ell\left(T_0, p_\ell\right)+\left(p_0-p_\ell\right) \frac{\partial \mu_\ell}{\partial p}\left(T_0, p_\ell\right) \\
& = \mu_\ell\left(T_0, p_\ell\right)+\left(p_0-p_\ell\right) v_\ell\left(T_0\right).
\end{aligned}
$$
On the right hand side, we recognize two terms appearing in the expression of $\Delta \Phi$ in (\ref{eq:Phi}), so that we can write 
$$
\Delta \Phi=n\left[\mu_\ell\left(T_0, p_0\right)-\mu_v\left(T_0, p_0\right)\right]+ F_s.
$$
This is the expression used in Section~\ref{sec:SingleDropDescription} (where it was adapted to also account for the substrate energy). It follows that the energetic gain $\Delta \Phi$ due to the phase change is the difference in chemical potentials, $\mu_\ell\left(T_0, p_0\right)-\mu_v\left(T_0, p_0\right)$, both taken at the reference pressure $p_0$, i.e. as if the surface were flat. This somewhat counterintuitive result arises from the fact that the liquid-vapor system is coupled to a reservoir at pressure $p_0$. Finding the energy barrier $\Delta \Phi^*$ amounts to finding the extremum of $\Delta \Phi$. At this extremum, corresponding to the (unstable) equilibrium, one naturally satisfies the conditions (\ref{eq:KelvinLaplace}). Finally, describing the vapour phase as an ideal gas, the thermodynamic potential is written under the form:
\begin{equation}\label{eq:deltaPhi}
\Delta \Phi=- n \Lambda k_B T+ F_s \quad {\rm with} \quad \Lambda=\frac{v_\ell  (p_{\rm sat}-p_0)}{k_BT}+\ln\left( \rho \big/ \rho_{\mathrm{sat}} \right),
\end{equation}
Apart close to a critical point, the term $v_\ell  (p_{\rm sat}-p_0)/(k_BT)$ is usually negligible.
}
%
%\textcolor{red}{Jacco: the factor $\Lambda$ is measured with respect to $p_{\rm sat}$, so we need to evaluate $\Delta \mu$ with respect to $p_{\rm sat}$. So, in the above expression, we make the approximation $p_0=p_{\rm sat}$. We do the same when solving the diffusion problem, i.e. we evaluate $\dot n$ based on $\rho_{\rm sat}$ at the drop. So it is consistent, but I think there IS an approximation made here. I think the more complete expression for $\Delta \mu$ reads: 
%%
%\begin{equation}
%\Delta \mu = \mu_\ell(p_0,T)-\mu_v(p_0,T) = 
%\mu_\ell(p_0,T) - \mu_\ell(p_{\rm sat},T) - \left( \mu_v(p_0,T) - \mu_v(p_{\rm sat},T\right) \equiv - \Lambda k_B T = , 
%\end{equation}
%% 
%Note that if we invoke the Kelvin-equation (appendix), which is allowed AT the critical nucleus, we find $\Delta \mu^* = -v_\ell \gamma \kappa$, which is in agreement with Elliot paper. So the result in geometric terms is simple, but in terms of $\Lambda$, we perhaps made an approximation? Note that $k_BT/v_\ell = 10^8$ Pa, so that the correction $p_0 - p_{\rm sat}$ is truly negligible. }

\section{Classical nucleation theory}\label{app:CNT}
We recall here the main results of the classical nucleation theory, necessary to understand the model hypotheses. Drops of type $A_n$ are clusters of $n$ molecules deposited on the substrate, which can grow, merge or break up. Their surface density is noted $\psi_n$. A single water molecule is a monomer $A_1^v$ in the vapor phase. We consider only binary processes that involve a monomer, \textit{i.e.} the aggregation of one vapor molecule at the surface of a drop containing $n$ water molecules. We first only focus on reversible growth, one molecule at a time, represented by the chemical balance equations
\begin{align}
 {A_1}^v &\stackrel{\beta_1}{\underset{\alpha_1}{\rightleftharpoons}} A_1 \nonumber\\
A_1 + {A_1}^v &\stackrel{\beta_2}{\underset{\alpha_2}{\rightleftharpoons}} A_2 \nonumber\\
&\, \vdots \nonumber \\
A_{n-1} + {A_1}^v &\stackrel{\beta_{n}}{\underset{\alpha_{n}}{\rightleftharpoons}} A_{n}\label{eq:rev_reactions}\\
&\, \vdots \nonumber
\end{align}
$\beta_{n}$ is the step-forward reaction rate from size $n-1$ to size $n$ while $\alpha_{n}$ the step-backward reaction rate. The number of particles follows as:
\begin{equation}
\frac{\mathrm d \psi_n}{\mathrm d t} = J_{n} - J_{n+1}\quad{\rm with}\quad J_{n} = \beta_n \psi_{n-1} - \alpha_{n} \psi_{n}
\label{eq:revgrowthstat}
\end{equation}
The vapor deposition rate $\beta_n$ onto a drop containing $n$ molecules has a contribution from the collisions from molecule in the surrounding air $\sim n^{2/3}$ and a contribution from the diffusion of molecules along the surface, proportional to $\sim n^{1/3}$. Introducing the cross-over value $n_c$ between these two regimes, we write the law under the generic form:
 $\beta_n=(n^{2/3}+n_c^{1/3} n^{1/3})\mathcal{B}$, where $\mathcal{B}$ is proportional to $\rho$. The adsorption stage can be written under the same form as eq.~(\ref{eq:revgrowthstat}), introducing a fictive density $\psi_0$ proportional to $\rho$. The equilibrium concentration obey Boltzmann law:
\begin{equation}
\psi_n^0 = \psi_0 \exp{\frac{-\Delta \Phi_n}{k_B T}}
\end{equation}
which allows one, using detailed balance, to express $\alpha_n$ as:
\begin{equation}
\alpha_{n} = \beta_n \frac{\psi_{n-1}^0}{\psi_{n}^0} =\beta_n \exp{\frac{\Delta \Phi_{n}-\Delta \Phi_{n-1}}{k_B T}}
\end{equation}
We therefore get the equation governing the flux of creation of nano-droplets containing $n$ molecules:
\begin{equation}
 J_{n}= \beta_{n} \psi_{n-1}^0 \left( \frac{\psi_{n-1}}{\psi_{n-1}^0 } -\frac{\psi_{n}}{\psi_{n}^0 } \right)
\label{eq:revgrowthstat2}
\end{equation}
In the steady state, the flux obeys $J_n= J$ for all $n$, where $J$ is the number of drops formed per unit surface and unit time. Dividing equation (\ref{eq:revgrowthstat}) by $\beta_n \psi_{n-1}^0$, and summing from $n = 1$ to $n =\infty$, we get:
\begin{equation}
J = \left(\sum_{n=1}^{\infty} \frac{1}{\beta_n \psi_{n-1}^0}\right)^{-1}={\mathcal B} \psi_0 \left(\sum_{n=1}^{\infty} \frac{\exp{\frac{\Delta \Phi_{n-1}}{k_B T}}}{ n^{2/3}+n_c^{1/3} n^{1/3} }\right)^{-1}
\end{equation}
In the limit $\Delta \Phi^* \gg k_B T$, the sum tends to its first term:
\begin{equation}
J ={\mathcal B} \psi_0 (1+n_c^{1/3})
\end{equation} 
In the activated regime $\Delta \Phi^* \ll k_B T$, the sum can been approximated by an integral, leading to:
\begin{equation}
J \simeq \beta_{n^*} \psi_0 Z_* \exp\left(-\frac{\Delta \Phi^*}{k_B T}\right)
\end{equation}
where $Z_*$ is called the Zeldovich factor, which reads:
\begin{equation}
Z_* = \sqrt{\frac{-1}{2\pi k_B T} \left.\frac{\mathrm d^2 \Delta \Phi}{\mathrm d n^2}\right|_{n}}=\frac{\Lambda^2 }{\sqrt{3\pi \chi}}% = \sqrt{\frac{1}{9\pi k_B T}} \left(\frac{3\Omega}{4\pi}\right)^{2/3} \left(\frac{k_B T \ln{(1+S)}}{2\gamma \Omega}\right)^2
\end{equation}
In the following, the calculations will be performed for $n_c=0$, for which the flux on nanodrops takes the final form, parametrized by $\chi$ and the multiplicative factor ${\mathcal J}$:
\begin{equation}
J \simeq {\mathcal J} \left(\frac{\rho}{\rho_{\rm sat}}\right)^2 \exp\left(- \frac{\chi}{\Lambda^2}\right)
\label{eq:rateenergyGap}
\end{equation}

\section{Electrostatic analogy}\label{app:electrostatics}

\begin{figure}
\centering
\includegraphics[width=1\textwidth]{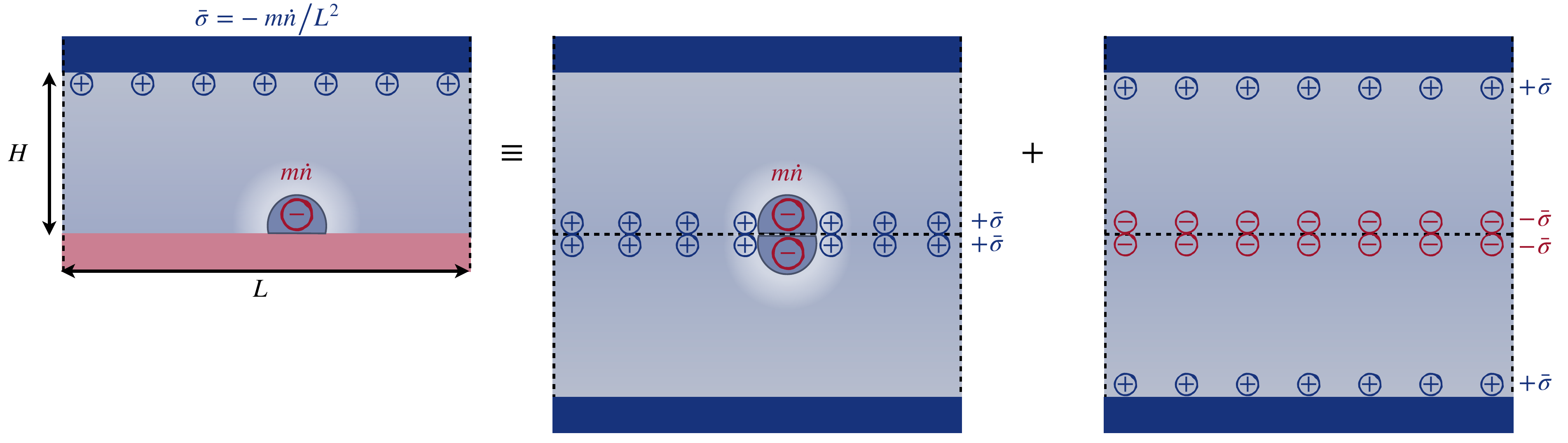}
\caption{Analogy with the fields created by a 2D periodic array of charge $\dot n$. The neutrality condition requires that the array produces an uniform backgound charge in $z=0$, compensating for the discrete charges carried by the drops. The non-accumulation condition requires that this uniform background charge in $z=0$ is compensated by the injection in $z=H$. }
\label{fig:ElectrostaticAnalogy}
\end{figure}

The far field drop problem is analogous to point charges in electrostatics, the charge here being proportional to the growth rate $\dot n_i$. In terms of electrostatics, the charge at the bottom ($z=0$) $\dot n_i$ (Fig~\ref{fig:ElectrostaticAnalogy}) is then balanced by a homogeneous surface charge density $-\dot n_i/L^2$ on the ceiling ($z=H$). In order to impose the boundary condition on the solid located at $z=0$, we introduce mirror charges. We decompose the problem as schematized in figure~\ref{fig:ElectrostaticAnalogy} into the contribution of the sinking drops in $z=0$, counterbalanced by an uniform background charge $-2\dot n/L^2$ in $z=0$ and the contribution of a double capacitor for which the plates carry an opposite charge $\pm \dot n/L^2$. We write formally:
\begin{equation}
\mathcal{G}(\mathbf r)=\sum_{i,j} \frac{1}{|\vec r-i L \vec e_x-j L \vec e_y|}+\mathcal{G}_g
%\label{eq:outer_asymptotics}
\end{equation}
where $i$ and $j$ label the periodic copies of the drops. The gauge value $\mathcal{G}_g$ is found using the condition: $\int \mathcal{G} d^2x=0$ over the ceiling. For later reference it is useful to split of the $1/r$ singularity according to: $\tilde{\mathcal G}(\vec r)=\mathcal G(\vec r) - |\vec r|^{-1}$ where $\tilde{\mathcal G}$ is regular, with a well-defined finite value $\tilde{\mathcal G}(\vec 0)$. The value is determined numerically in real space. 

The Green's function $\mathcal G$ captures not only the effect of each condensed drop, but also the uniform flux that feeds the drop from the ceiling and the boundary condition on the floor. The Green function $\mathcal G$ vanishes on the ceiling $z=H$. It is normalised such that a single drop gives a singularity $\mathcal G=1/|\vec r|$ and obey:
\begin{equation}\label{eq:poissonG}
\nabla^2 \mathcal G = - 2 \pi \sigma.
\end{equation}
with
\begin{equation}
\sigma(\vec r) = \sum_{p,q} 2 \delta (x-pL)\delta (y-qL)\delta(z) - \frac{1}{L^2} \left[\delta(z-H) + \delta(z+H) \right].
\end{equation}
Here we used the convention that a single drop represents two charge units: the true drop, associated with a sink $\dot n$ and the mirror drop with another sing $\dot n$, the latter of which of course does not contribute to the growth of the drop. We split the charge according to $\sigma=\sigma_1+\sigma_2$, with 
\begin{equation}
\sigma_1(\vec r) =\sum_{p,q} 2 \delta (x-pL)\delta (y-qL)\delta(z) - \frac{ 2}{L^2} \delta(z),
\end{equation}
where $(p,q)$ are integers used to scan the bi-periodic array in the $(x,y)$-plane. The residual charge reads:
\begin{equation}
\sigma_2(\vec r) = \frac{1}{L^2} \left[ 2\delta(z) - \delta(z-H) -\delta(z+H) \right].
\end{equation}
The latter has a simple solution $\mathcal G_2 =2\pi (H-|z|)/L^2$. The Fourier transform of the Poisson equation (eq.\ref{eq:poissonG}) gives $|\mathbf{K}|^2 \hat {\mathcal G}_1(\mathbf{K})=- 2\pi \hat \sigma_1(\mathbf{K})$
introducing the 3D-wavevector $\mathbf{K}=(k_x,k_y,k_z)$. The inverse Fourier-transform in $z$ can be performed, which gives:
\begin{equation}
\hat{\mathcal G}_1(\mathbf{k}, z)=\pi \frac{\hat \sigma_1(\mathbf{k})}{|\mathbf{k}|} e^{-|\mathbf{k}||z|}
\end{equation}
where $\mathbf{k}=\left(k_x, k_y\right)$ is the 2D-wavevector. For the discrete charges, $\dot n$ distributed along a periodic array in the $(x,y)$-plane, the wavevectors along the $(x,y)$ directions are quantified $k_x=2\pi p /L $, $k_y=2\pi q /L $ and the norm of the 2D-wavenumber is $|\mathbf{k}|=\dfrac{2\pi}{L} \sqrt{p^2+q^2}$. 
We thus get:
\begin{equation}
\mathcal G_1= \frac{1}{L}\sum_{p^2+q^2 \neq 0} \frac{1}{ \sqrt{p^2+q^2}} \cos \left( 2 \pi \left(\frac{ px+qy}{ L}\right) \right) \exp{\left(-2\pi \sqrt{p^2+q^2}\frac{z}{L}\right)}
\label{eq:GreenFunction3D}
\end{equation}

The total Green's function thus reads
\begin{eqnarray}
{\mathcal G}(\vec r) = \frac{2\pi}{L^2}(H-|z|)
+\sum_{p^2+q^2 \neq 0} \frac{1}{L \sqrt{p^2+q^2}} \cos \left( 2 \pi \left(\frac{ px+qy}{ L}\right) \right) \exp{\left(-2\pi \sqrt{p^2+q^2}\frac{z}{L}\right)}
\end{eqnarray}

\end{appendix}
\bibliography{bibmainPRF}

\providecommand{\noopsort}[1]{}\providecommand{\singleletter}[1]{#1}%
\begin{thebibliography}{66}%
\makeatletter
\providecommand \@ifxundefined [1]{%
 \@ifx{#1\undefined}
}%
\providecommand \@ifnum [1]{%
 \ifnum #1\expandafter \@firstoftwo
 \else \expandafter \@secondoftwo
 \fi
}%
\providecommand \@ifx [1]{%
 \ifx #1\expandafter \@firstoftwo
 \else \expandafter \@secondoftwo
 \fi
}%
\providecommand \natexlab [1]{#1}%
\providecommand \enquote  [1]{``#1''}%
\providecommand \bibnamefont  [1]{#1}%
\providecommand \bibfnamefont [1]{#1}%
\providecommand \citenamefont [1]{#1}%
\providecommand \href@noop [0]{\@secondoftwo}%
\providecommand \href [0]{\begingroup \@sanitize@url \@href}%
\providecommand \@href[1]{\@@startlink{#1}\@@href}%
\providecommand \@@href[1]{\endgroup#1\@@endlink}%
\providecommand \@sanitize@url [0]{\catcode `\\12\catcode `\$12\catcode
  `\&12\catcode `\#12\catcode `\^12\catcode `\_12\catcode `\%12\relax}%
\providecommand \@@startlink[1]{}%
\providecommand \@@endlink[0]{}%
\providecommand \url  [0]{\begingroup\@sanitize@url \@url }%
\providecommand \@url [1]{\endgroup\@href {#1}{\urlprefix }}%
\providecommand \urlprefix  [0]{URL }%
\providecommand \Eprint [0]{\href }%
\providecommand \doibase [0]{http://dx.doi.org/}%
\providecommand \selectlanguage [0]{\@gobble}%
\providecommand \bibinfo  [0]{\@secondoftwo}%
\providecommand \bibfield  [0]{\@secondoftwo}%
\providecommand \translation [1]{[#1]}%
\providecommand \BibitemOpen [0]{}%
\providecommand \bibitemStop [0]{}%
\providecommand \bibitemNoStop [0]{.\EOS\space}%
\providecommand \EOS [0]{\spacefactor3000\relax}%
\providecommand \BibitemShut  [1]{\csname bibitem#1\endcsname}%
\let\auto@bib@innerbib\@empty
%</preamble>
\bibitem [{\citenamefont {Rayleigh}(1911)}]{Rayleigh1911}%
  \BibitemOpen
  \bibfield  {author} {\bibinfo {author} {\bibfnamefont {L.}~\bibnamefont
  {Rayleigh}},\ }\bibfield  {title} {\enquote {\bibinfo {title} {{Breath
  Figures}},}\ }\href {\doibase 10.1259/jrs.1911.0067} {\bibfield  {journal}
  {\bibinfo  {journal} {J. R{\"{o}}ntgen Soc.}\ }\textbf {\bibinfo {volume}
  {7}},\ \bibinfo {pages} {126--126} (\bibinfo {year} {1911})}\BibitemShut
  {NoStop}%
\bibitem [{\citenamefont {Baker}(1922)}]{baker1922lxv}%
  \BibitemOpen
  \bibfield  {author} {\bibinfo {author} {\bibfnamefont {TJ}~\bibnamefont
  {Baker}},\ }\bibfield  {title} {\enquote {\bibinfo {title} {Lxv. breath
  figures},}\ }\href@noop {} {\bibfield  {journal} {\bibinfo  {journal} {The
  London, Edinburgh, and Dublin Philosophical Magazine and Journal of Science}\
  }\textbf {\bibinfo {volume} {44}},\ \bibinfo {pages} {752--765} (\bibinfo
  {year} {1922})}\BibitemShut {NoStop}%
\bibitem [{\citenamefont {Kashchiev}(2000)}]{kashchiev2000nucleation}%
  \BibitemOpen
  \bibfield  {author} {\bibinfo {author} {\bibfnamefont {Dimo}\ \bibnamefont
  {Kashchiev}},\ }\href@noop {} {\emph {\bibinfo {title} {Nucleation}}}\
  (\bibinfo  {publisher} {Elsevier},\ \bibinfo {year} {2000})\BibitemShut
  {NoStop}%
\bibitem [{\citenamefont {Beysens}(2022)}]{beysens2022physics}%
  \BibitemOpen
  \bibfield  {author} {\bibinfo {author} {\bibfnamefont {Daniel}\ \bibnamefont
  {Beysens}},\ }\href@noop {} {\emph {\bibinfo {title} {The physics of dew,
  breath figures and dropwise condensation}}},\ Vol.\ \bibinfo {volume} {994}\
  (\bibinfo  {publisher} {Springer},\ \bibinfo {year} {2022})\BibitemShut
  {NoStop}%
\bibitem [{\citenamefont {Knobler}\ \emph {et~al.}(1991)\citenamefont
  {Knobler}, \citenamefont {Steyer}, \citenamefont {Guenoun},\ and\
  \citenamefont {Fritter}}]{Knobler1991}%
  \BibitemOpen
  \bibfield  {author} {\bibinfo {author} {\bibfnamefont {C.M}\ \bibnamefont
  {Knobler}}, \bibinfo {author} {\bibfnamefont {A}~\bibnamefont {Steyer}},
  \bibinfo {author} {\bibfnamefont {P}~\bibnamefont {Guenoun}}, \ and\ \bibinfo
  {author} {\bibfnamefont {D}~\bibnamefont {Fritter}},\ }\bibfield  {title}
  {\enquote {\bibinfo {title} {{How Does Dew Form?}}}\ }\href {\doibase
  10.1080/01411599108206932} {\bibfield  {journal} {\bibinfo  {journal} {Phase
  Transit.}\ }\textbf {\bibinfo {volume} {31}},\ \bibinfo {pages} {219--246}
  (\bibinfo {year} {1991})}\BibitemShut {NoStop}%
\bibitem [{\citenamefont {Varanasi}\ \emph {et~al.}(2009)\citenamefont
  {Varanasi}, \citenamefont {Hsu}, \citenamefont {Bhate}, \citenamefont
  {Yang},\ and\ \citenamefont {Deng}}]{Varanasi2009}%
  \BibitemOpen
  \bibfield  {author} {\bibinfo {author} {\bibfnamefont {Kripa~K.}\
  \bibnamefont {Varanasi}}, \bibinfo {author} {\bibfnamefont {Ming}\
  \bibnamefont {Hsu}}, \bibinfo {author} {\bibfnamefont {Nitin}\ \bibnamefont
  {Bhate}}, \bibinfo {author} {\bibfnamefont {Wensha}\ \bibnamefont {Yang}}, \
  and\ \bibinfo {author} {\bibfnamefont {Tao}\ \bibnamefont {Deng}},\
  }\bibfield  {title} {\enquote {\bibinfo {title} {{Spatial control in the
  heterogeneous nucleation of water}},}\ }\href {\doibase 10.1063/1.3200951}
  {\bibfield  {journal} {\bibinfo  {journal} {Appl. Phys. Lett.}\ }\textbf
  {\bibinfo {volume} {95}},\ \bibinfo {pages} {2007--2010} (\bibinfo {year}
  {2009})}\BibitemShut {NoStop}%
\bibitem [{\citenamefont {Sikarwar}\ \emph {et~al.}(2011)\citenamefont
  {Sikarwar}, \citenamefont {Battoo}, \citenamefont {Khandekar},\ and\
  \citenamefont {Muralidhar}}]{Sikarwar2011}%
  \BibitemOpen
  \bibfield  {author} {\bibinfo {author} {\bibfnamefont {Basant~Singh}\
  \bibnamefont {Sikarwar}}, \bibinfo {author} {\bibfnamefont {Nirmal~Kumar}\
  \bibnamefont {Battoo}}, \bibinfo {author} {\bibfnamefont {Sameer}\
  \bibnamefont {Khandekar}}, \ and\ \bibinfo {author} {\bibfnamefont
  {K.}~\bibnamefont {Muralidhar}},\ }\bibfield  {title} {\enquote {\bibinfo
  {title} {{Dropwise condensation underneath chemically textured surfaces:
  Simulation and experiments}},}\ }\href {\doibase 10.1115/1.4002396}
  {\bibfield  {journal} {\bibinfo  {journal} {J. Heat Transf.}\ }\textbf
  {\bibinfo {volume} {133}},\ \bibinfo {pages} {021006} (\bibinfo {year}
  {2011})}\BibitemShut {NoStop}%
\bibitem [{\citenamefont {Lopez}\ \emph {et~al.}(1993)\citenamefont {Lopez},
  \citenamefont {Biebuyck}, \citenamefont {Frisbie},\ and\ \citenamefont
  {Whitesides}}]{Lopez1993}%
  \BibitemOpen
  \bibfield  {author} {\bibinfo {author} {\bibfnamefont {Gabriel~P}\
  \bibnamefont {Lopez}}, \bibinfo {author} {\bibfnamefont {Hans~a}\
  \bibnamefont {Biebuyck}}, \bibinfo {author} {\bibfnamefont {C~Daniel}\
  \bibnamefont {Frisbie}}, \ and\ \bibinfo {author} {\bibfnamefont {George~M}\
  \bibnamefont {Whitesides}},\ }\bibfield  {title} {\enquote {\bibinfo {title}
  {{REPORTS Imaging of Features}},}\ }\href@noop {} {\bibfield  {journal}
  {\bibinfo  {journal} {Science}\ }\textbf {\bibinfo {volume} {260}},\ \bibinfo
  {pages} {647--649} (\bibinfo {year} {1993})}\BibitemShut {NoStop}%
\bibitem [{\citenamefont {Enright}\ \emph {et~al.}(2012)\citenamefont
  {Enright}, \citenamefont {Miljkovic}, \citenamefont {Al-Obeidi},
  \citenamefont {Thompson},\ and\ \citenamefont {Wang}}]{Enright2012}%
  \BibitemOpen
  \bibfield  {author} {\bibinfo {author} {\bibfnamefont {Ryan}\ \bibnamefont
  {Enright}}, \bibinfo {author} {\bibfnamefont {Nenad}\ \bibnamefont
  {Miljkovic}}, \bibinfo {author} {\bibfnamefont {Ahmed}\ \bibnamefont
  {Al-Obeidi}}, \bibinfo {author} {\bibfnamefont {Carl~V.}\ \bibnamefont
  {Thompson}}, \ and\ \bibinfo {author} {\bibfnamefont {Evelyn~N.}\
  \bibnamefont {Wang}},\ }\bibfield  {title} {\enquote {\bibinfo {title}
  {{Condensation on superhydrophobic surfaces: The role of local energy
  barriers and structure length scale}},}\ }\href {\doibase 10.1021/la302599n}
  {\bibfield  {journal} {\bibinfo  {journal} {Langmuir}\ }\textbf {\bibinfo
  {volume} {28}},\ \bibinfo {pages} {14424--14432} (\bibinfo {year}
  {2012})}\BibitemShut {NoStop}%
\bibitem [{\citenamefont {Gelderblom}\ \emph {et~al.}(2011)\citenamefont
  {Gelderblom}, \citenamefont {Mar{\'{i}}n}, \citenamefont {Nair},
  \citenamefont {van Houselt}, \citenamefont {Lefferts}, \citenamefont
  {Snoeijer},\ and\ \citenamefont {Lohse}}]{Gelderblom2011}%
  \BibitemOpen
  \bibfield  {author} {\bibinfo {author} {\bibfnamefont {Hanneke}\ \bibnamefont
  {Gelderblom}}, \bibinfo {author} {\bibfnamefont {{\'{A}}lvaro~G.}\
  \bibnamefont {Mar{\'{i}}n}}, \bibinfo {author} {\bibfnamefont {Hrudya}\
  \bibnamefont {Nair}}, \bibinfo {author} {\bibfnamefont {Arie}\ \bibnamefont
  {van Houselt}}, \bibinfo {author} {\bibfnamefont {Leon}\ \bibnamefont
  {Lefferts}}, \bibinfo {author} {\bibfnamefont {Jacco~H.}\ \bibnamefont
  {Snoeijer}}, \ and\ \bibinfo {author} {\bibfnamefont {Detlef}\ \bibnamefont
  {Lohse}},\ }\bibfield  {title} {\enquote {\bibinfo {title} {{How water
  droplets evaporate on a superhydrophobic substrate}},}\ }\href {\doibase
  10.1103/PhysRevE.83.026306} {\bibfield  {journal} {\bibinfo  {journal} {Phys.
  Rev. E}\ }\textbf {\bibinfo {volume} {83}},\ \bibinfo {pages} {026306}
  (\bibinfo {year} {2011})}\BibitemShut {NoStop}%
\bibitem [{\citenamefont {Popov}(2005)}]{Popov2005}%
  \BibitemOpen
  \bibfield  {author} {\bibinfo {author} {\bibfnamefont {Yuri~O}\ \bibnamefont
  {Popov}},\ }\bibfield  {title} {\enquote {\bibinfo {title} {Evaporative
  deposition patterns: spatial dimensions of the deposit},}\ }\href@noop {}
  {\bibfield  {journal} {\bibinfo  {journal} {Phys. Rev. E}\ }\textbf {\bibinfo
  {volume} {71}},\ \bibinfo {pages} {036313} (\bibinfo {year}
  {2005})}\BibitemShut {NoStop}%
\bibitem [{\citenamefont {Nath}\ \emph {et~al.}(2018)\citenamefont {Nath},
  \citenamefont {Bisbano}, \citenamefont {Yue},\ and\ \citenamefont
  {Boreyko}}]{nath2018duelling}%
  \BibitemOpen
  \bibfield  {author} {\bibinfo {author} {\bibfnamefont {Saurabh}\ \bibnamefont
  {Nath}}, \bibinfo {author} {\bibfnamefont {Caitlin~E}\ \bibnamefont
  {Bisbano}}, \bibinfo {author} {\bibfnamefont {Pengtao}\ \bibnamefont {Yue}},
  \ and\ \bibinfo {author} {\bibfnamefont {Jonathan~B}\ \bibnamefont
  {Boreyko}},\ }\bibfield  {title} {\enquote {\bibinfo {title} {Duelling dry
  zones around hygroscopic droplets},}\ }\href {\doibase 10.1017/jfm.2018.579}
  {\bibfield  {journal} {\bibinfo  {journal} {Journal of Fluid Mechanics}\
  }\textbf {\bibinfo {volume} {853}},\ \bibinfo {pages} {601--620} (\bibinfo
  {year} {2018})}\BibitemShut {NoStop}%
\bibitem [{\citenamefont {Guadarrama-Cetina}\ \emph {et~al.}(2014)\citenamefont
  {Guadarrama-Cetina}, \citenamefont {Narhe}, \citenamefont {Beysens},\ and\
  \citenamefont {Gonzalez-Vinas}}]{guadarrama2014droplet}%
  \BibitemOpen
  \bibfield  {author} {\bibinfo {author} {\bibfnamefont {J}~\bibnamefont
  {Guadarrama-Cetina}}, \bibinfo {author} {\bibfnamefont {RD}~\bibnamefont
  {Narhe}}, \bibinfo {author} {\bibfnamefont {DA}~\bibnamefont {Beysens}}, \
  and\ \bibinfo {author} {\bibfnamefont {W}~\bibnamefont {Gonzalez-Vinas}},\
  }\bibfield  {title} {\enquote {\bibinfo {title} {Droplet pattern and
  condensation gradient around a humidity sink},}\ }\href {\doibase
  10.1103/PhysRevE.89.012402} {\bibfield  {journal} {\bibinfo  {journal}
  {Physical Review E}\ }\textbf {\bibinfo {volume} {89}},\ \bibinfo {pages}
  {012402} (\bibinfo {year} {2014})}\BibitemShut {NoStop}%
\bibitem [{\citenamefont {Schafle}\ \emph {et~al.}(2003)\citenamefont
  {Schafle}, \citenamefont {Leiderer},\ and\ \citenamefont
  {Bechinger}}]{schafle2003subpattern}%
  \BibitemOpen
  \bibfield  {author} {\bibinfo {author} {\bibfnamefont {C}~\bibnamefont
  {Schafle}}, \bibinfo {author} {\bibfnamefont {P}~\bibnamefont {Leiderer}}, \
  and\ \bibinfo {author} {\bibfnamefont {C}~\bibnamefont {Bechinger}},\
  }\bibfield  {title} {\enquote {\bibinfo {title} {Subpattern formation during
  condensation processes on structured substrates},}\ }\href {\doibase
  10.1209/epl/i2003-00540-7} {\bibfield  {journal} {\bibinfo  {journal}
  {Europhysics Letters}\ }\textbf {\bibinfo {volume} {63}},\ \bibinfo {pages}
  {394--400} (\bibinfo {year} {2003})}\BibitemShut {NoStop}%
\bibitem [{\citenamefont {Yu}\ \emph {et~al.}(2021)\citenamefont {Yu},
  \citenamefont {Dorao},\ and\ \citenamefont {Fernandino}}]{yu2021droplet}%
  \BibitemOpen
  \bibfield  {author} {\bibinfo {author} {\bibfnamefont {Xiongjiang}\
  \bibnamefont {Yu}}, \bibinfo {author} {\bibfnamefont {Carlos~Alberto}\
  \bibnamefont {Dorao}}, \ and\ \bibinfo {author} {\bibfnamefont {Maria}\
  \bibnamefont {Fernandino}},\ }\bibfield  {title} {\enquote {\bibinfo {title}
  {Droplet evaporation during dropwise condensation due to deposited volatile
  organic compounds},}\ }\href {\doibase 10.1063/5.0056005} {\bibfield
  {journal} {\bibinfo  {journal} {AIP Advances}\ }\textbf {\bibinfo {volume}
  {11}} (\bibinfo {year} {2021}),\ 10.1063/5.0056005}\BibitemShut {NoStop}%
\bibitem [{\citenamefont {Beysens}(2006)}]{Beysens2006}%
  \BibitemOpen
  \bibfield  {author} {\bibinfo {author} {\bibfnamefont {Daniel}\ \bibnamefont
  {Beysens}},\ }\bibfield  {title} {\enquote {\bibinfo {title} {{Dew nucleation
  and growth}},}\ }\href {\doibase 10.1016/j.crhy.2006.10.020} {\bibfield
  {journal} {\bibinfo  {journal} {Comptes Rendus Phys.}\ }\textbf {\bibinfo
  {volume} {7}},\ \bibinfo {pages} {1082--1100} (\bibinfo {year}
  {2006})}\BibitemShut {NoStop}%
\bibitem [{\citenamefont {Briscoe}\ and\ \citenamefont
  {Galvin}(1991)}]{Briscoe1991}%
  \BibitemOpen
  \bibfield  {author} {\bibinfo {author} {\bibfnamefont {B.~J.}\ \bibnamefont
  {Briscoe}}\ and\ \bibinfo {author} {\bibfnamefont {K.~P.}\ \bibnamefont
  {Galvin}},\ }\bibfield  {title} {\enquote {\bibinfo {title} {{An experimental
  study of the growth of breath figures}},}\ }\href {\doibase
  10.1016/0166-6622(91)80126-9} {\bibfield  {journal} {\bibinfo  {journal}
  {Colloids Surf.}\ }\textbf {\bibinfo {volume} {56}},\ \bibinfo {pages}
  {263--278} (\bibinfo {year} {1991})}\BibitemShut {NoStop}%
\bibitem [{\citenamefont {Sokuler}\ \emph
  {et~al.}(2010{\natexlab{a}})\citenamefont {Sokuler}, \citenamefont
  {Auernhammer}, \citenamefont {Liu}, \citenamefont {Bonaccurso},\ and\
  \citenamefont {Butt}}]{Sokuler2010interspacing}%
  \BibitemOpen
  \bibfield  {author} {\bibinfo {author} {\bibfnamefont {M}~\bibnamefont
  {Sokuler}}, \bibinfo {author} {\bibfnamefont {G{\"u}nter~K}\ \bibnamefont
  {Auernhammer}}, \bibinfo {author} {\bibfnamefont {CJ}~\bibnamefont {Liu}},
  \bibinfo {author} {\bibfnamefont {Elmar}\ \bibnamefont {Bonaccurso}}, \ and\
  \bibinfo {author} {\bibfnamefont {Hans-J{\"u}rgen}\ \bibnamefont {Butt}},\
  }\bibfield  {title} {\enquote {\bibinfo {title} {Dynamics of condensation and
  evaporation: Effect of inter-drop spacing},}\ }\href@noop {} {\bibfield
  {journal} {\bibinfo  {journal} {EPL}\ }\textbf {\bibinfo {volume} {89}},\
  \bibinfo {pages} {36004} (\bibinfo {year} {2010}{\natexlab{a}})}\BibitemShut
  {NoStop}%
\bibitem [{\citenamefont {Picknett}\ and\ \citenamefont
  {Bexon}(1977)}]{picknett1977evaporation}%
  \BibitemOpen
  \bibfield  {author} {\bibinfo {author} {\bibfnamefont {RG}~\bibnamefont
  {Picknett}}\ and\ \bibinfo {author} {\bibfnamefont {R}~\bibnamefont
  {Bexon}},\ }\bibfield  {title} {\enquote {\bibinfo {title} {The evaporation
  of sessile or pendant drops in still air},}\ }\href@noop {} {\bibfield
  {journal} {\bibinfo  {journal} {Journal of colloid and Interface Science}\
  }\textbf {\bibinfo {volume} {61}},\ \bibinfo {pages} {336--350} (\bibinfo
  {year} {1977})}\BibitemShut {NoStop}%
\bibitem [{\citenamefont {Bintein}\ \emph {et~al.}(2019)\citenamefont
  {Bintein}, \citenamefont {Lhuissier}, \citenamefont {Mongruel}, \citenamefont
  {Royon},\ and\ \citenamefont {Beysens}}]{Bintein2019}%
  \BibitemOpen
  \bibfield  {author} {\bibinfo {author} {\bibfnamefont {Pierre~Brice}\
  \bibnamefont {Bintein}}, \bibinfo {author} {\bibfnamefont {Henri}\
  \bibnamefont {Lhuissier}}, \bibinfo {author} {\bibfnamefont {Anne}\
  \bibnamefont {Mongruel}}, \bibinfo {author} {\bibfnamefont {Laurent}\
  \bibnamefont {Royon}}, \ and\ \bibinfo {author} {\bibfnamefont {Daniel}\
  \bibnamefont {Beysens}},\ }\bibfield  {title} {\enquote {\bibinfo {title}
  {{Grooves Accelerate Dew Shedding}},}\ }\href {\doibase
  10.1103/PhysRevLett.122.098005} {\bibfield  {journal} {\bibinfo  {journal}
  {Phys. Rev. Lett.}\ }\textbf {\bibinfo {volume} {122}},\ \bibinfo {pages}
  {98005} (\bibinfo {year} {2019})}\BibitemShut {NoStop}%
\bibitem [{\citenamefont {Trosseille}\ \emph {et~al.}(2019)\citenamefont
  {Trosseille}, \citenamefont {Mongruel}, \citenamefont {Royon}, \citenamefont
  {Medici},\ and\ \citenamefont {Beysens}}]{Trosseille2019}%
  \BibitemOpen
  \bibfield  {author} {\bibinfo {author} {\bibfnamefont {Joachim}\ \bibnamefont
  {Trosseille}}, \bibinfo {author} {\bibfnamefont {Anne}\ \bibnamefont
  {Mongruel}}, \bibinfo {author} {\bibfnamefont {Laurent}\ \bibnamefont
  {Royon}}, \bibinfo {author} {\bibfnamefont {Marie~Gabrielle}\ \bibnamefont
  {Medici}}, \ and\ \bibinfo {author} {\bibfnamefont {Daniel}\ \bibnamefont
  {Beysens}},\ }\bibfield  {title} {\enquote {\bibinfo {title}
  {{Roughness-enhanced collection of condensed droplets}},}\ }\href {\doibase
  10.1140/epje/i2019-11905-9} {\bibfield  {journal} {\bibinfo  {journal}
  {EPJE}\ }\textbf {\bibinfo {volume} {42}} (\bibinfo {year} {2019}),\
  10.1140/epje/i2019-11905-9}\BibitemShut {NoStop}%
\bibitem [{\citenamefont {Viovy}\ \emph {et~al.}(1988)\citenamefont {Viovy},
  \citenamefont {Beysens},\ and\ \citenamefont {Knobler}}]{Viovy1988scaling}%
  \BibitemOpen
  \bibfield  {author} {\bibinfo {author} {\bibfnamefont {Jean~Louis}\
  \bibnamefont {Viovy}}, \bibinfo {author} {\bibfnamefont {Daniel}\
  \bibnamefont {Beysens}}, \ and\ \bibinfo {author} {\bibfnamefont {Charles~M}\
  \bibnamefont {Knobler}},\ }\bibfield  {title} {\enquote {\bibinfo {title}
  {Scaling description for the growth of condensation patterns on surfaces},}\
  }\href@noop {} {\bibfield  {journal} {\bibinfo  {journal} {Phys. Rev. A}\
  }\textbf {\bibinfo {volume} {37}},\ \bibinfo {pages} {4965} (\bibinfo {year}
  {1988})}\BibitemShut {NoStop}%
\bibitem [{\citenamefont {Blaschke}\ \emph {et~al.}(2012)\citenamefont
  {Blaschke}, \citenamefont {Lapp}, \citenamefont {Hof},\ and\ \citenamefont
  {Vollmer}}]{Blaschke2012}%
  \BibitemOpen
  \bibfield  {author} {\bibinfo {author} {\bibfnamefont {Johannes}\
  \bibnamefont {Blaschke}}, \bibinfo {author} {\bibfnamefont {Tobias}\
  \bibnamefont {Lapp}}, \bibinfo {author} {\bibfnamefont {Bj{\"{o}}rn}\
  \bibnamefont {Hof}}, \ and\ \bibinfo {author} {\bibfnamefont {J{\"{u}}rgen}\
  \bibnamefont {Vollmer}},\ }\bibfield  {title} {\enquote {\bibinfo {title}
  {{Breath figures: Nucleation, growth, coalescence, and the size distribution
  of droplets}},}\ }\href {\doibase 10.1103/PhysRevLett.109.068701} {\bibfield
  {journal} {\bibinfo  {journal} {Phys. Rev. Lett.}\ }\textbf {\bibinfo
  {volume} {109}},\ \bibinfo {pages} {3--6} (\bibinfo {year}
  {2012})}\BibitemShut {NoStop}%
\bibitem [{\citenamefont {Stricker}\ \emph {et~al.}(2022)\citenamefont
  {Stricker}, \citenamefont {Grillo}, \citenamefont {Marquez}, \citenamefont
  {Panzarasa}, \citenamefont {Smith-Mannschott},\ and\ \citenamefont
  {Vollmer}}]{Stricker2022universality}%
  \BibitemOpen
  \bibfield  {author} {\bibinfo {author} {\bibfnamefont {Laura}\ \bibnamefont
  {Stricker}}, \bibinfo {author} {\bibfnamefont {Fabio}\ \bibnamefont
  {Grillo}}, \bibinfo {author} {\bibfnamefont {EA}~\bibnamefont {Marquez}},
  \bibinfo {author} {\bibfnamefont {Guido}\ \bibnamefont {Panzarasa}}, \bibinfo
  {author} {\bibfnamefont {Katrina}\ \bibnamefont {Smith-Mannschott}}, \ and\
  \bibinfo {author} {\bibfnamefont {J{\"u}rgen}\ \bibnamefont {Vollmer}},\
  }\bibfield  {title} {\enquote {\bibinfo {title} {Universality of breath
  figures on two-dimensional surfaces: An experimental study},}\ }\href
  {\doibase 10.1103/PhysRevResearch.4.L012019} {\bibfield  {journal} {\bibinfo
  {journal} {Phys. Rev. Res.}\ }\textbf {\bibinfo {volume} {4}},\ \bibinfo
  {pages} {L012019} (\bibinfo {year} {2022})}\BibitemShut {NoStop}%
\bibitem [{\citenamefont {Haderbache}\ \emph {et~al.}(1998)\citenamefont
  {Haderbache}, \citenamefont {Garrigos}, \citenamefont {Kofman}, \citenamefont
  {S{\o}ndergard},\ and\ \citenamefont {Cheyssac}}]{haderbache1998numerical}%
  \BibitemOpen
  \bibfield  {author} {\bibinfo {author} {\bibfnamefont {L}~\bibnamefont
  {Haderbache}}, \bibinfo {author} {\bibfnamefont {R}~\bibnamefont {Garrigos}},
  \bibinfo {author} {\bibfnamefont {R}~\bibnamefont {Kofman}}, \bibinfo
  {author} {\bibfnamefont {E}~\bibnamefont {S{\o}ndergard}}, \ and\ \bibinfo
  {author} {\bibfnamefont {P}~\bibnamefont {Cheyssac}},\ }\bibfield  {title}
  {\enquote {\bibinfo {title} {Numerical and experimental investigations of the
  size ordering of nanocrystals},}\ }\href@noop {} {\bibfield  {journal}
  {\bibinfo  {journal} {Surf. Sci.}\ }\textbf {\bibinfo {volume} {410}},\
  \bibinfo {pages} {L748--L756} (\bibinfo {year} {1998})}\BibitemShut {NoStop}%
\bibitem [{\citenamefont {Nikolayev}\ \emph {et~al.}(1996)\citenamefont
  {Nikolayev}, \citenamefont {Beysens}, \citenamefont {Gioda}, \citenamefont
  {Milimouka}, \citenamefont {Katiushin},\ and\ \citenamefont
  {Morel}}]{nikolayev1996water}%
  \BibitemOpen
  \bibfield  {author} {\bibinfo {author} {\bibfnamefont {VS}~\bibnamefont
  {Nikolayev}}, \bibinfo {author} {\bibfnamefont {D}~\bibnamefont {Beysens}},
  \bibinfo {author} {\bibfnamefont {Alain}\ \bibnamefont {Gioda}}, \bibinfo
  {author} {\bibfnamefont {I}~\bibnamefont {Milimouka}}, \bibinfo {author}
  {\bibfnamefont {E}~\bibnamefont {Katiushin}}, \ and\ \bibinfo {author}
  {\bibfnamefont {J-P}\ \bibnamefont {Morel}},\ }\bibfield  {title} {\enquote
  {\bibinfo {title} {Water recovery from dew},}\ }\href@noop {} {\bibfield
  {journal} {\bibinfo  {journal} {J. Hydrol}\ }\textbf {\bibinfo {volume}
  {182}},\ \bibinfo {pages} {19--35} (\bibinfo {year} {1996})}\BibitemShut
  {NoStop}%
\bibitem [{\citenamefont {Liu}\ \emph {et~al.}(2022)\citenamefont {Liu},
  \citenamefont {Beysens},\ and\ \citenamefont {Bourouina}}]{liu2022water}%
  \BibitemOpen
  \bibfield  {author} {\bibinfo {author} {\bibfnamefont {Xiaoyi}\ \bibnamefont
  {Liu}}, \bibinfo {author} {\bibfnamefont {Daniel}\ \bibnamefont {Beysens}}, \
  and\ \bibinfo {author} {\bibfnamefont {Tarik}\ \bibnamefont {Bourouina}},\
  }\bibfield  {title} {\enquote {\bibinfo {title} {Water harvesting from air:
  Current passive approaches and outlook},}\ }\href@noop {} {\bibfield
  {journal} {\bibinfo  {journal} {ACS Mater. Lett.}\ }\textbf {\bibinfo
  {volume} {4}},\ \bibinfo {pages} {1003--1024} (\bibinfo {year}
  {2022})}\BibitemShut {NoStop}%
\bibitem [{\citenamefont {Parker}\ and\ \citenamefont
  {Lawrence}(2001)}]{parker2001water}%
  \BibitemOpen
  \bibfield  {author} {\bibinfo {author} {\bibfnamefont {Andrew~R}\
  \bibnamefont {Parker}}\ and\ \bibinfo {author} {\bibfnamefont {Chris~R}\
  \bibnamefont {Lawrence}},\ }\bibfield  {title} {\enquote {\bibinfo {title}
  {Water capture by a desert beetle},}\ }\href@noop {} {\bibfield  {journal}
  {\bibinfo  {journal} {Nature}\ }\textbf {\bibinfo {volume} {414}},\ \bibinfo
  {pages} {33--34} (\bibinfo {year} {2001})}\BibitemShut {NoStop}%
\bibitem [{\citenamefont {Munn{\'e}-Bosch}\ and\ \citenamefont
  {Alegre}(1999)}]{munne1999role}%
  \BibitemOpen
  \bibfield  {author} {\bibinfo {author} {\bibfnamefont {Sergi}\ \bibnamefont
  {Munn{\'e}-Bosch}}\ and\ \bibinfo {author} {\bibfnamefont {Leonor}\
  \bibnamefont {Alegre}},\ }\bibfield  {title} {\enquote {\bibinfo {title}
  {Role of dew on the recovery of water-stressed melissa officinalis l.
  plants},}\ }\href@noop {} {\bibfield  {journal} {\bibinfo  {journal} {J.
  Plant Physiol.}\ }\textbf {\bibinfo {volume} {154}},\ \bibinfo {pages}
  {759--766} (\bibinfo {year} {1999})}\BibitemShut {NoStop}%
\bibitem [{\citenamefont {Hill}\ \emph {et~al.}(2015)\citenamefont {Hill},
  \citenamefont {Dawson}, \citenamefont {Shelef},\ and\ \citenamefont
  {Rachmilevitch}}]{hill2015role}%
  \BibitemOpen
  \bibfield  {author} {\bibinfo {author} {\bibfnamefont {Amber~J}\ \bibnamefont
  {Hill}}, \bibinfo {author} {\bibfnamefont {Todd~E}\ \bibnamefont {Dawson}},
  \bibinfo {author} {\bibfnamefont {Oren}\ \bibnamefont {Shelef}}, \ and\
  \bibinfo {author} {\bibfnamefont {Shimon}\ \bibnamefont {Rachmilevitch}},\
  }\bibfield  {title} {\enquote {\bibinfo {title} {The role of dew in negev
  desert plants},}\ }\href@noop {} {\bibfield  {journal} {\bibinfo  {journal}
  {Oecologia}\ }\textbf {\bibinfo {volume} {178}},\ \bibinfo {pages} {317--327}
  (\bibinfo {year} {2015})}\BibitemShut {NoStop}%
\bibitem [{\citenamefont {Bortolin}\ \emph {et~al.}(2022)\citenamefont
  {Bortolin}, \citenamefont {Tancon},\ and\ \citenamefont
  {Del~Col}}]{bortolin2022heat}%
  \BibitemOpen
  \bibfield  {author} {\bibinfo {author} {\bibfnamefont {Stefano}\ \bibnamefont
  {Bortolin}}, \bibinfo {author} {\bibfnamefont {Marco}\ \bibnamefont
  {Tancon}}, \ and\ \bibinfo {author} {\bibfnamefont {Davide}\ \bibnamefont
  {Del~Col}},\ }\enquote {\bibinfo {title} {Heat transfer enhancement during
  dropwise condensation over wettability-controlled surfaces},}\ in\ \href
  {\doibase 10.1007/978-3-030-82992-6_3} {\emph {\bibinfo {booktitle} {The
  Surface Wettability Effect on Phase Change}}},\ \bibinfo {editor} {edited by\
  \bibinfo {editor} {\bibfnamefont {Marco}\ \bibnamefont {Marengo}}\ and\
  \bibinfo {editor} {\bibfnamefont {Joel}\ \bibnamefont {De~Coninck}}}\
  (\bibinfo  {publisher} {Springer},\ \bibinfo {year} {2022})\ pp.\ \bibinfo
  {pages} {29--67}\BibitemShut {NoStop}%
\bibitem [{\citenamefont {Khawaji}\ \emph {et~al.}(2008)\citenamefont
  {Khawaji}, \citenamefont {Kutubkhanah},\ and\ \citenamefont
  {Wie}}]{khawaji2008advances}%
  \BibitemOpen
  \bibfield  {author} {\bibinfo {author} {\bibfnamefont {Akili~D}\ \bibnamefont
  {Khawaji}}, \bibinfo {author} {\bibfnamefont {Ibrahim~K}\ \bibnamefont
  {Kutubkhanah}}, \ and\ \bibinfo {author} {\bibfnamefont {Jong-Mihn}\
  \bibnamefont {Wie}},\ }\bibfield  {title} {\enquote {\bibinfo {title}
  {Advances in seawater desalination technologies},}\ }\href@noop {} {\bibfield
   {journal} {\bibinfo  {journal} {Desalination}\ }\textbf {\bibinfo {volume}
  {221}},\ \bibinfo {pages} {47--69} (\bibinfo {year} {2008})}\BibitemShut
  {NoStop}%
\bibitem [{\citenamefont {Guha}\ \emph {et~al.}(2017)\citenamefont {Guha},
  \citenamefont {Anand},\ and\ \citenamefont {Varanasi}}]{Guha2017}%
  \BibitemOpen
  \bibfield  {author} {\bibinfo {author} {\bibfnamefont {Ingrid~F.}\
  \bibnamefont {Guha}}, \bibinfo {author} {\bibfnamefont {Sushant}\
  \bibnamefont {Anand}}, \ and\ \bibinfo {author} {\bibfnamefont {Kripa~K.}\
  \bibnamefont {Varanasi}},\ }\bibfield  {title} {\enquote {\bibinfo {title}
  {{Creating nanoscale emulsions using condensation}},}\ }\href {\doibase
  10.1038/s41467-017-01420-8} {\bibfield  {journal} {\bibinfo  {journal} {Nat.
  Commun.}\ }\textbf {\bibinfo {volume} {8}},\ \bibinfo {pages} {1--6}
  (\bibinfo {year} {2017})}\BibitemShut {NoStop}%
\bibitem [{\citenamefont {Goodling}\ \emph {et~al.}(2019)\citenamefont
  {Goodling}, \citenamefont {Nagelberg}, \citenamefont {Kaehr}, \citenamefont
  {Meredith}, \citenamefont {Cheon}, \citenamefont {Saunders}, \citenamefont
  {Kolle},\ and\ \citenamefont {Zarzar}}]{goodling2019colouration}%
  \BibitemOpen
  \bibfield  {author} {\bibinfo {author} {\bibfnamefont {Amy~E}\ \bibnamefont
  {Goodling}}, \bibinfo {author} {\bibfnamefont {Sara}\ \bibnamefont
  {Nagelberg}}, \bibinfo {author} {\bibfnamefont {Bryan}\ \bibnamefont
  {Kaehr}}, \bibinfo {author} {\bibfnamefont {Caleb~H}\ \bibnamefont
  {Meredith}}, \bibinfo {author} {\bibfnamefont {Seong~Ik}\ \bibnamefont
  {Cheon}}, \bibinfo {author} {\bibfnamefont {Ashley~P}\ \bibnamefont
  {Saunders}}, \bibinfo {author} {\bibfnamefont {Mathias}\ \bibnamefont
  {Kolle}}, \ and\ \bibinfo {author} {\bibfnamefont {Lauren~D}\ \bibnamefont
  {Zarzar}},\ }\bibfield  {title} {\enquote {\bibinfo {title} {Colouration by
  total internal reflection and interference at microscale concave
  interfaces},}\ }\href@noop {} {\bibfield  {journal} {\bibinfo  {journal}
  {Nature}\ }\textbf {\bibinfo {volume} {566}},\ \bibinfo {pages} {523--527}
  (\bibinfo {year} {2019})}\BibitemShut {NoStop}%
\bibitem [{\citenamefont {B{\"o}ker}\ \emph {et~al.}(2004)\citenamefont
  {B{\"o}ker}, \citenamefont {Lin}, \citenamefont {Chiapperini}, \citenamefont
  {Horowitz}, \citenamefont {Thompson}, \citenamefont {Carreon}, \citenamefont
  {Xu}, \citenamefont {Abetz}, \citenamefont {Skaff}, \citenamefont {Dinsmore}
  \emph {et~al.}}]{boker2004hierarchical}%
  \BibitemOpen
  \bibfield  {author} {\bibinfo {author} {\bibfnamefont {Alexander}\
  \bibnamefont {B{\"o}ker}}, \bibinfo {author} {\bibfnamefont {Yao}\
  \bibnamefont {Lin}}, \bibinfo {author} {\bibfnamefont {Kristen}\ \bibnamefont
  {Chiapperini}}, \bibinfo {author} {\bibfnamefont {Reina}\ \bibnamefont
  {Horowitz}}, \bibinfo {author} {\bibfnamefont {Mike}\ \bibnamefont
  {Thompson}}, \bibinfo {author} {\bibfnamefont {Vincent}\ \bibnamefont
  {Carreon}}, \bibinfo {author} {\bibfnamefont {Ting}\ \bibnamefont {Xu}},
  \bibinfo {author} {\bibfnamefont {Clarissa}\ \bibnamefont {Abetz}}, \bibinfo
  {author} {\bibfnamefont {Habib}\ \bibnamefont {Skaff}}, \bibinfo {author}
  {\bibfnamefont {AD}~\bibnamefont {Dinsmore}},  \emph {et~al.},\ }\bibfield
  {title} {\enquote {\bibinfo {title} {Hierarchical nanoparticle assemblies
  formed by decorating breath figures},}\ }\href@noop {} {\bibfield  {journal}
  {\bibinfo  {journal} {Nat. Mater.}\ }\textbf {\bibinfo {volume} {3}},\
  \bibinfo {pages} {302--306} (\bibinfo {year} {2004})}\BibitemShut {NoStop}%
\bibitem [{\citenamefont {Zhang}\ \emph {et~al.}(2015)\citenamefont {Zhang},
  \citenamefont {Bai},\ and\ \citenamefont {Li}}]{zhang2015breath}%
  \BibitemOpen
  \bibfield  {author} {\bibinfo {author} {\bibfnamefont {Aijuan}\ \bibnamefont
  {Zhang}}, \bibinfo {author} {\bibfnamefont {Hua}\ \bibnamefont {Bai}}, \ and\
  \bibinfo {author} {\bibfnamefont {Lei}\ \bibnamefont {Li}},\ }\bibfield
  {title} {\enquote {\bibinfo {title} {Breath figure: a nature-inspired
  preparation method for ordered porous films},}\ }\href@noop {} {\bibfield
  {journal} {\bibinfo  {journal} {Chem. Rev.}\ }\textbf {\bibinfo {volume}
  {115}},\ \bibinfo {pages} {9801--9868} (\bibinfo {year} {2015})}\BibitemShut
  {NoStop}%
\bibitem [{\citenamefont {Park}\ \emph {et~al.}(2016)\citenamefont {Park},
  \citenamefont {Kim}, \citenamefont {Grinthal}, \citenamefont {He},
  \citenamefont {Fox}, \citenamefont {Weaver},\ and\ \citenamefont
  {Aizenberg}}]{Park2016}%
  \BibitemOpen
  \bibfield  {author} {\bibinfo {author} {\bibfnamefont {Kyoo~Chul}\
  \bibnamefont {Park}}, \bibinfo {author} {\bibfnamefont {Philseok}\
  \bibnamefont {Kim}}, \bibinfo {author} {\bibfnamefont {Alison}\ \bibnamefont
  {Grinthal}}, \bibinfo {author} {\bibfnamefont {Neil}\ \bibnamefont {He}},
  \bibinfo {author} {\bibfnamefont {David}\ \bibnamefont {Fox}}, \bibinfo
  {author} {\bibfnamefont {James~C.}\ \bibnamefont {Weaver}}, \ and\ \bibinfo
  {author} {\bibfnamefont {Joanna}\ \bibnamefont {Aizenberg}},\ }\bibfield
  {title} {\enquote {\bibinfo {title} {{Condensation on slippery asymmetric
  bumps}},}\ }\href {\doibase 10.1038/nature16956} {\bibfield  {journal}
  {\bibinfo  {journal} {Nature}\ }\textbf {\bibinfo {volume} {531}},\ \bibinfo
  {pages} {78--82} (\bibinfo {year} {2016})}\BibitemShut {NoStop}%
\bibitem [{\citenamefont {Zhao}\ and\ \citenamefont
  {Beysens}(1995)}]{Zhao1995}%
  \BibitemOpen
  \bibfield  {author} {\bibinfo {author} {\bibfnamefont {Hong}\ \bibnamefont
  {Zhao}}\ and\ \bibinfo {author} {\bibfnamefont {Daniel}\ \bibnamefont
  {Beysens}},\ }\bibfield  {title} {\enquote {\bibinfo {title} {{From Droplet
  Growth to Film Growth on a Heterogeneous Surface: Condensation Associated
  with a Wettability Gradient}},}\ }\href {\doibase 10.1021/la00002a045}
  {\bibfield  {journal} {\bibinfo  {journal} {Langmuir}\ }\textbf {\bibinfo
  {volume} {11}},\ \bibinfo {pages} {627--634} (\bibinfo {year}
  {1995})}\BibitemShut {NoStop}%
\bibitem [{\citenamefont {Baratian}\ \emph {et~al.}(2018)\citenamefont
  {Baratian}, \citenamefont {Dey}, \citenamefont {Hoek}, \citenamefont {{Van
  Den Ende}},\ and\ \citenamefont {Mugele}}]{Baratian2018}%
  \BibitemOpen
  \bibfield  {author} {\bibinfo {author} {\bibfnamefont {Davood}\ \bibnamefont
  {Baratian}}, \bibinfo {author} {\bibfnamefont {Ranabir}\ \bibnamefont {Dey}},
  \bibinfo {author} {\bibfnamefont {Harmen}\ \bibnamefont {Hoek}}, \bibinfo
  {author} {\bibfnamefont {Dirk}\ \bibnamefont {{Van Den Ende}}}, \ and\
  \bibinfo {author} {\bibfnamefont {Frieder}\ \bibnamefont {Mugele}},\
  }\bibfield  {title} {\enquote {\bibinfo {title} {{Breath Figures under
  Electrowetting: Electrically Controlled Evolution of Drop Condensation
  Patterns}},}\ }\href {\doibase 10.1103/PhysRevLett.120.214502} {\bibfield
  {journal} {\bibinfo  {journal} {Phys. Rev. Lett.}\ }\textbf {\bibinfo
  {volume} {120}},\ \bibinfo {pages} {214502} (\bibinfo {year}
  {2018})}\BibitemShut {NoStop}%
\bibitem [{\citenamefont {Bouillant}\ \emph {et~al.}(2024)\citenamefont
  {Bouillant}, \citenamefont {Henkel}, \citenamefont {Thiele}, \citenamefont
  {Andreotti},\ and\ \citenamefont {Snoeijer}}]{PRL_companion}%
  \BibitemOpen
  \bibfield  {author} {\bibinfo {author} {\bibfnamefont {Ambre}\ \bibnamefont
  {Bouillant}}, \bibinfo {author} {\bibfnamefont {Christopher}\ \bibnamefont
  {Henkel}}, \bibinfo {author} {\bibfnamefont {Uwe}\ \bibnamefont {Thiele}},
  \bibinfo {author} {\bibfnamefont {Bruno}\ \bibnamefont {Andreotti}}, \ and\
  \bibinfo {author} {\bibfnamefont {Jacco~H}\ \bibnamefont {Snoeijer}},\
  }\bibfield  {title} {\enquote {\bibinfo {title} {Soft condensation},}\
  }\href@noop {} {\bibfield  {journal} {\bibinfo  {journal} {Submitted to Phys.
  Rev. Lett. as a companion paper}\ } (\bibinfo {year} {2024})}\BibitemShut
  {NoStop}%
\bibitem [{\citenamefont {Kelton}\ and\ \citenamefont
  {Greer}(2010)}]{kelton2010nucleation}%
  \BibitemOpen
  \bibfield  {author} {\bibinfo {author} {\bibfnamefont {Ken}\ \bibnamefont
  {Kelton}}\ and\ \bibinfo {author} {\bibfnamefont {Alan}\ \bibnamefont
  {Greer}},\ }\href@noop {} {\emph {\bibinfo {title} {Nucleation in condensed
  matter: applications in materials \& biology}}}\ (\bibinfo  {publisher}
  {Elsevier},\ \bibinfo {year} {2010})\BibitemShut {NoStop}%
\bibitem [{\citenamefont {Eslami}\ and\ \citenamefont
  {Elliott}(2011)}]{eslami2011thermodynamic}%
  \BibitemOpen
  \bibfield  {author} {\bibinfo {author} {\bibfnamefont {Fatemeh}\ \bibnamefont
  {Eslami}}\ and\ \bibinfo {author} {\bibfnamefont {Janet~AW}\ \bibnamefont
  {Elliott}},\ }\bibfield  {title} {\enquote {\bibinfo {title} {Thermodynamic
  investigation of the barrier for heterogeneous nucleation on a fluid surface
  in comparison with a rigid surface},}\ }\href@noop {} {\bibfield  {journal}
  {\bibinfo  {journal} {The Journal of Physical Chemistry B}\ }\textbf
  {\bibinfo {volume} {115}},\ \bibinfo {pages} {10646--10653} (\bibinfo {year}
  {2011})}\BibitemShut {NoStop}%
\bibitem [{\citenamefont {Diu}\ \emph {et~al.}(2007)\citenamefont {Diu},
  \citenamefont {Guthmann},\ and\ \citenamefont
  {Lederer}}]{diu2007thermodynamique}%
  \BibitemOpen
  \bibfield  {author} {\bibinfo {author} {\bibfnamefont {Bernard}\ \bibnamefont
  {Diu}}, \bibinfo {author} {\bibfnamefont {Claudine}\ \bibnamefont
  {Guthmann}}, \ and\ \bibinfo {author} {\bibfnamefont {Danielle}\ \bibnamefont
  {Lederer}},\ }\href@noop {} {\emph {\bibinfo {title} {Thermodynamique}}}\
  (\bibinfo  {publisher} {Editions Hermann},\ \bibinfo {year}
  {2007})\BibitemShut {NoStop}%
\bibitem [{\citenamefont {Wu}(1996)}]{wu1996nucleation}%
  \BibitemOpen
  \bibfield  {author} {\bibinfo {author} {\bibfnamefont {David~T}\ \bibnamefont
  {Wu}},\ }\bibfield  {title} {\enquote {\bibinfo {title} {Nucleation
  theory},}\ }in\ \href@noop {} {\emph {\bibinfo {booktitle} {Solid State
  Physics}}},\ Vol.~\bibinfo {volume} {50}\ (\bibinfo  {publisher} {Elsevier},\
  \bibinfo {year} {1996})\ pp.\ \bibinfo {pages} {37--187}\BibitemShut
  {NoStop}%
\bibitem [{\citenamefont {Rogers}\ \emph {et~al.}(1988)\citenamefont {Rogers},
  \citenamefont {Elder},\ and\ \citenamefont {Desai}}]{Rogers1988}%
  \BibitemOpen
  \bibfield  {author} {\bibinfo {author} {\bibfnamefont {T.~M.}\ \bibnamefont
  {Rogers}}, \bibinfo {author} {\bibfnamefont {K.~R.}\ \bibnamefont {Elder}}, \
  and\ \bibinfo {author} {\bibfnamefont {Rashmi~C.}\ \bibnamefont {Desai}},\
  }\bibfield  {title} {\enquote {\bibinfo {title} {{Droplet growth and
  coarsening during heterogeneous vapor condensation}},}\ }\href {\doibase
  10.1103/PhysRevA.38.5303} {\bibfield  {journal} {\bibinfo  {journal} {Phys.
  Rev. A}\ }\textbf {\bibinfo {volume} {38}},\ \bibinfo {pages} {5303--5309}
  (\bibinfo {year} {1988})}\BibitemShut {NoStop}%
\bibitem [{\citenamefont {Fritter}\ \emph {et~al.}(1988)\citenamefont
  {Fritter}, \citenamefont {Knobler}, \citenamefont {Roux},\ and\ \citenamefont
  {Beysens}}]{Fritter1988}%
  \BibitemOpen
  \bibfield  {author} {\bibinfo {author} {\bibfnamefont {Daniela}\ \bibnamefont
  {Fritter}}, \bibinfo {author} {\bibfnamefont {Charles~M.}\ \bibnamefont
  {Knobler}}, \bibinfo {author} {\bibfnamefont {Didier}\ \bibnamefont {Roux}},
  \ and\ \bibinfo {author} {\bibfnamefont {Daniel}\ \bibnamefont {Beysens}},\
  }\bibfield  {title} {\enquote {\bibinfo {title} {{Computer simulations of the
  growth of breath figures}},}\ }\href {\doibase 10.1007/BF01011659} {\bibfield
   {journal} {\bibinfo  {journal} {J Stat. Phys.}\ }\textbf {\bibinfo {volume}
  {52}},\ \bibinfo {pages} {1447--1459} (\bibinfo {year} {1988})}\BibitemShut
  {NoStop}%
\bibitem [{\citenamefont {Derrida}\ \emph {et~al.}(1991)\citenamefont
  {Derrida}, \citenamefont {Godreche},\ and\ \citenamefont
  {Yekutieli}}]{Derrida1991}%
  \BibitemOpen
  \bibfield  {author} {\bibinfo {author} {\bibfnamefont {B}~\bibnamefont
  {Derrida}}, \bibinfo {author} {\bibfnamefont {C}~\bibnamefont {Godreche}}, \
  and\ \bibinfo {author} {\bibfnamefont {I}~\bibnamefont {Yekutieli}},\
  }\bibfield  {title} {\enquote {\bibinfo {title} {Scale-invariant regimes in
  one-dimensional models of growing and coalescing droplets},}\ }\href@noop {}
  {\bibfield  {journal} {\bibinfo  {journal} {Phys. Rev. A}\ }\textbf {\bibinfo
  {volume} {44}},\ \bibinfo {pages} {6241} (\bibinfo {year}
  {1991})}\BibitemShut {NoStop}%
\bibitem [{\citenamefont {Meakin}(1992)}]{Meakin1992}%
  \BibitemOpen
  \bibfield  {author} {\bibinfo {author} {\bibfnamefont {Paul}\ \bibnamefont
  {Meakin}},\ }\bibfield  {title} {\enquote {\bibinfo {title} {{Droplet
  deposition growth and coalescence}},}\ }\href {\doibase
  10.1088/0034-4885/55/2/002} {\bibfield  {journal} {\bibinfo  {journal} {Rep.
  Prog. Phys.}\ }\textbf {\bibinfo {volume} {55}},\ \bibinfo {pages} {157--240}
  (\bibinfo {year} {1992})}\BibitemShut {NoStop}%
\bibitem [{\citenamefont {Zhang}\ \emph {et~al.}(2020)\citenamefont {Zhang},
  \citenamefont {Mei}, \citenamefont {Botto},\ and\ \citenamefont
  {Yang}}]{Zhang2020}%
  \BibitemOpen
  \bibfield  {author} {\bibinfo {author} {\bibfnamefont {Ran}\ \bibnamefont
  {Zhang}}, \bibinfo {author} {\bibfnamefont {Ran~Andy}\ \bibnamefont {Mei}},
  \bibinfo {author} {\bibfnamefont {Lorenzo}\ \bibnamefont {Botto}}, \ and\
  \bibinfo {author} {\bibfnamefont {Zhongqiang}\ \bibnamefont {Yang}},\
  }\bibfield  {title} {\enquote {\bibinfo {title} {{Modified Voronoi Analysis
  of Spontaneous Formation of Interfacial Droplets on Immersed Oil-Solid
  Substrates}},}\ }\href {\doibase 10.1021/acs.langmuir.9b03806} {\bibfield
  {journal} {\bibinfo  {journal} {Langmuir}\ }\textbf {\bibinfo {volume}
  {36}},\ \bibinfo {pages} {5400--5407} (\bibinfo {year} {2020})}\BibitemShut
  {NoStop}%
\bibitem [{\citenamefont {Steyer}\ \emph {et~al.}(1990)\citenamefont {Steyer},
  \citenamefont {Guenoun}, \citenamefont {Beysens},\ and\ \citenamefont
  {Knobler}}]{Steyer1990}%
  \BibitemOpen
  \bibfield  {author} {\bibinfo {author} {\bibfnamefont {A.}~\bibnamefont
  {Steyer}}, \bibinfo {author} {\bibfnamefont {P.}~\bibnamefont {Guenoun}},
  \bibinfo {author} {\bibfnamefont {D.}~\bibnamefont {Beysens}}, \ and\
  \bibinfo {author} {\bibfnamefont {C.~M.}\ \bibnamefont {Knobler}},\
  }\bibfield  {title} {\enquote {\bibinfo {title} {{Two-dimensional ordering
  during droplet growth on a liquid surface}},}\ }\href {\doibase
  10.1103/PhysRevB.42.1086} {\bibfield  {journal} {\bibinfo  {journal} {Phys.
  Rev. B}\ }\textbf {\bibinfo {volume} {42}},\ \bibinfo {pages} {1086--1089}
  (\bibinfo {year} {1990})}\BibitemShut {NoStop}%
\bibitem [{\citenamefont {Steyer}\ \emph {et~al.}(1993)\citenamefont {Steyer},
  \citenamefont {Guenoun},\ and\ \citenamefont {Beysens}}]{Steyer1993}%
  \BibitemOpen
  \bibfield  {author} {\bibinfo {author} {\bibfnamefont {A}~\bibnamefont
  {Steyer}}, \bibinfo {author} {\bibfnamefont {P}~\bibnamefont {Guenoun}}, \
  and\ \bibinfo {author} {\bibfnamefont {D}~\bibnamefont {Beysens}},\
  }\bibfield  {title} {\enquote {\bibinfo {title} {Hexatic and fat-fractal
  structures for water droplets condensing on oil},}\ }\href@noop {} {\bibfield
   {journal} {\bibinfo  {journal} {Phys. Rev. E}\ }\textbf {\bibinfo {volume}
  {48}},\ \bibinfo {pages} {428} (\bibinfo {year} {1993})}\BibitemShut
  {NoStop}%
\bibitem [{\citenamefont {Nepomnyashchy}\ \emph {et~al.}(2006)\citenamefont
  {Nepomnyashchy}, \citenamefont {Golovin}, \citenamefont {Tikhomirova},\ and\
  \citenamefont {Volpert}}]{Nepomnyashchy2006}%
  \BibitemOpen
  \bibfield  {author} {\bibinfo {author} {\bibfnamefont {A.}~\bibnamefont
  {Nepomnyashchy}}, \bibinfo {author} {\bibfnamefont {A.}~\bibnamefont
  {Golovin}}, \bibinfo {author} {\bibfnamefont {A.}~\bibnamefont
  {Tikhomirova}}, \ and\ \bibinfo {author} {\bibfnamefont {V.}~\bibnamefont
  {Volpert}},\ }\bibfield  {title} {\enquote {\bibinfo {title} {{Nucleation and
  growth of droplets at a liquid-gas interface}},}\ }\href {\doibase
  10.1103/PhysRevE.74.021605} {\bibfield  {journal} {\bibinfo  {journal} {Phys.
  Rev. E}\ }\textbf {\bibinfo {volume} {74}},\ \bibinfo {pages} {1--10}
  (\bibinfo {year} {2006})}\BibitemShut {NoStop}%
\bibitem [{\citenamefont {Anand}\ \emph {et~al.}(2015)\citenamefont {Anand},
  \citenamefont {Rykaczewski}, \citenamefont {Subramanyam}, \citenamefont
  {Beysens},\ and\ \citenamefont {Varanasi}}]{Anand2015}%
  \BibitemOpen
  \bibfield  {author} {\bibinfo {author} {\bibfnamefont {Sushant}\ \bibnamefont
  {Anand}}, \bibinfo {author} {\bibfnamefont {Konrad}\ \bibnamefont
  {Rykaczewski}}, \bibinfo {author} {\bibfnamefont {Srinivas~Bengaluru}\
  \bibnamefont {Subramanyam}}, \bibinfo {author} {\bibfnamefont {Daniel}\
  \bibnamefont {Beysens}}, \ and\ \bibinfo {author} {\bibfnamefont {Kripa~K.}\
  \bibnamefont {Varanasi}},\ }\bibfield  {title} {\enquote {\bibinfo {title}
  {{How droplets nucleate and grow on liquids and liquid impregnated
  surfaces}},}\ }\href {\doibase 10.1039/c4sm01424c} {\bibfield  {journal}
  {\bibinfo  {journal} {Soft Matter}\ }\textbf {\bibinfo {volume} {11}},\
  \bibinfo {pages} {69--80} (\bibinfo {year} {2015})}\BibitemShut {NoStop}%
\bibitem [{\citenamefont {Sharma}\ \emph {et~al.}(2022)\citenamefont {Sharma},
  \citenamefont {Milionis}, \citenamefont {Naga}, \citenamefont {Lam},
  \citenamefont {Rodriguez}, \citenamefont {{Del Ponte}}, \citenamefont
  {Negri}, \citenamefont {Raoul}, \citenamefont {D'Acunzi}, \citenamefont
  {Butt}, \citenamefont {Vollmer},\ and\ \citenamefont
  {Poulikakos}}]{Sharma2022}%
  \BibitemOpen
  \bibfield  {author} {\bibinfo {author} {\bibfnamefont {Chander~Shekhar}\
  \bibnamefont {Sharma}}, \bibinfo {author} {\bibfnamefont {Athanasios}\
  \bibnamefont {Milionis}}, \bibinfo {author} {\bibfnamefont {Abhinav}\
  \bibnamefont {Naga}}, \bibinfo {author} {\bibfnamefont {Cheuk Wing~Edmond}\
  \bibnamefont {Lam}}, \bibinfo {author} {\bibfnamefont {Gabriel}\ \bibnamefont
  {Rodriguez}}, \bibinfo {author} {\bibfnamefont {Marco~Francesco}\
  \bibnamefont {{Del Ponte}}}, \bibinfo {author} {\bibfnamefont {Valentina}\
  \bibnamefont {Negri}}, \bibinfo {author} {\bibfnamefont {Hopf}\ \bibnamefont
  {Raoul}}, \bibinfo {author} {\bibfnamefont {Maria}\ \bibnamefont {D'Acunzi}},
  \bibinfo {author} {\bibfnamefont {Hans~J{"{u}}rgen}\ \bibnamefont {Butt}},
  \bibinfo {author} {\bibfnamefont {Doris}\ \bibnamefont {Vollmer}}, \ and\
  \bibinfo {author} {\bibfnamefont {Dimos}\ \bibnamefont {Poulikakos}},\
  }\bibfield  {title} {\enquote {\bibinfo {title} {{Enhanced Condensation on
  Soft Materials through Bulk Lubricant Infusion}},}\ }\href {\doibase
  10.1002/adfm.202109633} {\bibfield  {journal} {\bibinfo  {journal} {Adv.
  Funct. Mater.}\ }\textbf {\bibinfo {volume} {32}},\ \bibinfo {pages}
  {202109633} (\bibinfo {year} {2022})}\BibitemShut {NoStop}%
\bibitem [{\citenamefont {Ge}\ \emph {et~al.}(2020)\citenamefont {Ge},
  \citenamefont {Raza}, \citenamefont {Li}, \citenamefont {Sett}, \citenamefont
  {Miljkovic},\ and\ \citenamefont {Zhang}}]{Ge2020}%
  \BibitemOpen
  \bibfield  {author} {\bibinfo {author} {\bibfnamefont {Qiaoyu}\ \bibnamefont
  {Ge}}, \bibinfo {author} {\bibfnamefont {Aikifa}\ \bibnamefont {Raza}},
  \bibinfo {author} {\bibfnamefont {Hongxia}\ \bibnamefont {Li}}, \bibinfo
  {author} {\bibfnamefont {Soumyadip}\ \bibnamefont {Sett}}, \bibinfo {author}
  {\bibfnamefont {Nenad}\ \bibnamefont {Miljkovic}}, \ and\ \bibinfo {author}
  {\bibfnamefont {Tiejun}\ \bibnamefont {Zhang}},\ }\bibfield  {title}
  {\enquote {\bibinfo {title} {{Condensation of Satellite Droplets on
  Lubricant-Cloaked Droplets}},}\ }\href {\doibase 10.1021/acsami.9b22417}
  {\bibfield  {journal} {\bibinfo  {journal} {ACS Appl. Mater. Interfaces}\
  }\textbf {\bibinfo {volume} {12}},\ \bibinfo {pages} {22246--22255} (\bibinfo
  {year} {2020})}\BibitemShut {NoStop}%
\bibitem [{\citenamefont {Sokuler}\ \emph
  {et~al.}(2010{\natexlab{b}})\citenamefont {Sokuler}, \citenamefont
  {Auernhammer}, \citenamefont {Roth}, \citenamefont {Liu}, \citenamefont
  {Bonacurrso},\ and\ \citenamefont {Butt}}]{Sokuler2010}%
  \BibitemOpen
  \bibfield  {author} {\bibinfo {author} {\bibfnamefont {Mordechai}\
  \bibnamefont {Sokuler}}, \bibinfo {author} {\bibfnamefont {G{\"{u}}nter~K.}\
  \bibnamefont {Auernhammer}}, \bibinfo {author} {\bibfnamefont {Marcel}\
  \bibnamefont {Roth}}, \bibinfo {author} {\bibfnamefont {Chuanjun}\
  \bibnamefont {Liu}}, \bibinfo {author} {\bibfnamefont {Elmar}\ \bibnamefont
  {Bonacurrso}}, \ and\ \bibinfo {author} {\bibfnamefont {Hans~J{\"{u}}rgen}\
  \bibnamefont {Butt}},\ }\bibfield  {title} {\enquote {\bibinfo {title} {{The
  softer the better: Fast condensation on soft surfaces}},}\ }\href {\doibase
  10.1021/la903996j} {\bibfield  {journal} {\bibinfo  {journal} {Langmuir}\
  }\textbf {\bibinfo {volume} {26}},\ \bibinfo {pages} {1544--1547} (\bibinfo
  {year} {2010}{\natexlab{b}})}\BibitemShut {NoStop}%
\bibitem [{\citenamefont {Villermaux}\ and\ \citenamefont
  {Innocenti}(1999)}]{Villermaux1999}%
  \BibitemOpen
  \bibfield  {author} {\bibinfo {author} {\bibfnamefont {Emmanuel}\
  \bibnamefont {Villermaux}}\ and\ \bibinfo {author} {\bibfnamefont {Claudia}\
  \bibnamefont {Innocenti}},\ }\bibfield  {title} {\enquote {\bibinfo {title}
  {On the geometry of turbulent mixing},}\ }\href {\doibase
  10.1017/S0022112099005674} {\bibfield  {journal} {\bibinfo  {journal} {J.
  Fluid Mech.}\ }\textbf {\bibinfo {volume} {393}},\ \bibinfo {pages}
  {123--147} (\bibinfo {year} {1999})}\BibitemShut {NoStop}%
\bibitem [{\citenamefont {Villermaux}(2019)}]{Villermaux2019}%
  \BibitemOpen
  \bibfield  {author} {\bibinfo {author} {\bibfnamefont {Emmanuel}\
  \bibnamefont {Villermaux}},\ }\bibfield  {title} {\enquote {\bibinfo {title}
  {Mixing versus stirring},}\ \ }(\bibinfo  {publisher} {Annual Reviews},\
  \bibinfo {year} {2019})\ pp.\ \bibinfo {pages} {245--273}\BibitemShut
  {NoStop}%
\bibitem [{\citenamefont {Family}\ and\ \citenamefont
  {Meakin}(1989)}]{family1989kinetics}%
  \BibitemOpen
  \bibfield  {author} {\bibinfo {author} {\bibfnamefont {Fereydoon}\
  \bibnamefont {Family}}\ and\ \bibinfo {author} {\bibfnamefont {Paul}\
  \bibnamefont {Meakin}},\ }\bibfield  {title} {\enquote {\bibinfo {title}
  {Kinetics of droplet growth processes: Simulations, theory, and
  experiments},}\ }\href@noop {} {\bibfield  {journal} {\bibinfo  {journal}
  {Phys. Rev. A}\ }\textbf {\bibinfo {volume} {40}},\ \bibinfo {pages} {3836}
  (\bibinfo {year} {1989})}\BibitemShut {NoStop}%
\bibitem [{\citenamefont {Katselas}\ \emph {et~al.}(2022)\citenamefont
  {Katselas}, \citenamefont {Parin},\ and\ \citenamefont
  {Neto}}]{Katselas2022}%
  \BibitemOpen
  \bibfield  {author} {\bibinfo {author} {\bibfnamefont {Anthony}\ \bibnamefont
  {Katselas}}, \bibinfo {author} {\bibfnamefont {Riccardo}\ \bibnamefont
  {Parin}}, \ and\ \bibinfo {author} {\bibfnamefont {Chiara}\ \bibnamefont
  {Neto}},\ }\bibfield  {title} {\enquote {\bibinfo {title} {Quantification of
  nucleation site density as a function of surface wettability on smooth
  surfaces},}\ }\href@noop {} {\bibfield  {journal} {\bibinfo  {journal} {Adv.
  Mater. Interfaces}\ }\textbf {\bibinfo {volume} {9}},\ \bibinfo {pages}
  {2200246} (\bibinfo {year} {2022})}\BibitemShut {NoStop}%
\bibitem [{\citenamefont {Rose}(2002)}]{Rose2002}%
  \BibitemOpen
  \bibfield  {author} {\bibinfo {author} {\bibfnamefont {J.~W.}\ \bibnamefont
  {Rose}},\ }\bibfield  {title} {\enquote {\bibinfo {title} {{Dropwise
  condensation theory and experiment: A review}},}\ }\href {\doibase
  10.1243/09576500260049034} {\bibfield  {journal} {\bibinfo  {journal} {Proc.
  Inst. Mech. Eng. Part A J. Power Energy}\ }\textbf {\bibinfo {volume}
  {216}},\ \bibinfo {pages} {115--128} (\bibinfo {year} {2002})}\BibitemShut
  {NoStop}%
\bibitem [{\citenamefont {Phadnis}\ and\ \citenamefont
  {Rykaczewski}(2017)}]{Phadnis2017}%
  \BibitemOpen
  \bibfield  {author} {\bibinfo {author} {\bibfnamefont {Akshay}\ \bibnamefont
  {Phadnis}}\ and\ \bibinfo {author} {\bibfnamefont {Konrad}\ \bibnamefont
  {Rykaczewski}},\ }\bibfield  {title} {\enquote {\bibinfo {title} {{Dropwise
  Condensation on Soft Hydrophobic Coatings}},}\ }\href {\doibase
  10.1021/acs.langmuir.7b03141} {\bibfield  {journal} {\bibinfo  {journal}
  {Langmuir}\ }\textbf {\bibinfo {volume} {33}},\ \bibinfo {pages}
  {12095--12101} (\bibinfo {year} {2017})}\BibitemShut {NoStop}%
\bibitem [{\citenamefont {Kolb}(1989)}]{Kolb1989}%
  \BibitemOpen
  \bibfield  {author} {\bibinfo {author} {\bibfnamefont {M.}~\bibnamefont
  {Kolb}},\ }\bibfield  {title} {\enquote {\bibinfo {title} {{Comment on
  Scaling of the droplet-size distribution in vapor-deposited thin films}},}\
  }\href {\doibase 10.1103/PhysRevLett.62.1699} {\bibfield  {journal} {\bibinfo
   {journal} {Phys. Rev. Lett.}\ }\textbf {\bibinfo {volume} {62}},\ \bibinfo
  {pages} {1699} (\bibinfo {year} {1989})}\BibitemShut {NoStop}%
\bibitem [{\citenamefont {Leach}\ \emph {et~al.}(2006)\citenamefont {Leach},
  \citenamefont {Stevens}, \citenamefont {Langford},\ and\ \citenamefont
  {Dickinson}}]{Leach2006}%
  \BibitemOpen
  \bibfield  {author} {\bibinfo {author} {\bibfnamefont {R.~N.}\ \bibnamefont
  {Leach}}, \bibinfo {author} {\bibfnamefont {F.}~\bibnamefont {Stevens}},
  \bibinfo {author} {\bibfnamefont {S.~C.}\ \bibnamefont {Langford}}, \ and\
  \bibinfo {author} {\bibfnamefont {J.~T.}\ \bibnamefont {Dickinson}},\
  }\bibfield  {title} {\enquote {\bibinfo {title} {{Dropwise condensation:
  Experiments and simulations of nucleation and growth of water drops in a
  cooling system}},}\ }\href {\doibase 10.1021/la061901+} {\bibfield  {journal}
  {\bibinfo  {journal} {Langmuir}\ }\textbf {\bibinfo {volume} {22}},\ \bibinfo
  {pages} {8864--8872} (\bibinfo {year} {2006})}\BibitemShut {NoStop}%
\bibitem [{\citenamefont {Lavielle}\ \emph {et~al.}(2023)\citenamefont
  {Lavielle}, \citenamefont {Beysens},\ and\ \citenamefont
  {Mongruel}}]{Lavielle2023}%
  \BibitemOpen
  \bibfield  {author} {\bibinfo {author} {\bibfnamefont {Nicolas}\ \bibnamefont
  {Lavielle}}, \bibinfo {author} {\bibfnamefont {Daniel}\ \bibnamefont
  {Beysens}}, \ and\ \bibinfo {author} {\bibfnamefont {Anne}\ \bibnamefont
  {Mongruel}},\ }\bibfield  {title} {\enquote {\bibinfo {title}
  {Nucleation-enhanced condensation and fast shedding on self-lubricated
  silicone organogels},}\ }\href {\doibase 10.1039/D3SM00365E} {\bibfield
  {journal} {\bibinfo  {journal} {Soft Matter}\ }\textbf {\bibinfo {volume}
  {19}},\ \bibinfo {pages} {4458--4469} (\bibinfo {year} {2023})}\BibitemShut
  {NoStop}%
\bibitem [{\citenamefont {Frenkel}(1955)}]{Frenkel}%
  \BibitemOpen
  \bibfield  {author} {\bibinfo {author} {\bibfnamefont {J.}~\bibnamefont
  {Frenkel}},\ }\href {https://cds.cern.ch/record/106808} {\emph {\bibinfo
  {title} {Kinetic theory of liquids}}}\ (\bibinfo  {publisher} {Dover
  Publications, Inc.},\ \bibinfo {year} {1955})\BibitemShut {NoStop}%
\end{thebibliography}%

\end{document}